\documentclass{emulateapj}

\usepackage{lineno} 
\usepackage{lscape}
\usepackage{footnote}
\usepackage{times}
\usepackage{graphicx}
\usepackage{verbatim} 
\usepackage{tabularx}
\usepackage{hhline}

\newcommand\T{\rule{0pt}{2.6ex}}
\newcommand\B{\rule[-1.2ex]{0pt}{0pt}}
\newcolumntype{Z}{>{\centering\arraybackslash}X}

\shorttitle{Multi-Zone Modeling of HESS J1825$-$137}
\shortauthors{Van Etten \& Romani}

\begin{document}

\title{Multi-Zone Modeling Of The Pulsar Wind Nebula HESS J1825$-$137}

\author{Adam Van Etten\altaffilmark{1} and Roger W. Romani\altaffilmark{1}}

\altaffiltext{1}{Department of Physics, Stanford University, Stanford, CA 94305}

\keywords{pulsars: individual (PSR J1826$-$1334) -- X-rays: general}

\begin{abstract}

The pulsar wind nebula associated with PSR J1826$-$1334, HESS J1825$-$137,
is a bright very high energy source with an angular extent of $\sim1^{\circ}$ and
spatially-resolved spectroscopic TeV measurements. The gamma-ray
spectral index is observed to soften with increasing distance from
the pulsar, likely the result of cooling losses as electrons
traverse the nebula.  
We describe
analysis of X-ray data of the extended nebula, as well as 3-D time-dependent
spectral energy distribution modeling, with emphasis on the spatial variations
within HESS J1825$-$137. The multi-wavelength data 
places significant constraints on electron injection, transport,
and cooling within the nebula.
The large size and high nebular energy budget imply a relatively rapid initial pulsar
spin period of $13\pm7$ ms and an age of $40\pm9$ kyr.
The relative fluxes of each VHE zone can be explained by advective particle transport
with a radially decreasing velocity profile with $v(r) \propto r^{-0.5}$. 
The evolution of the cooling break requires an evolving magnetic field which also decreases
radially from the pulsar, 
$B(r,t) \propto r^{-0.7} \, \dot{E}(t)^{1/2}$.
Detection of $10$ TeV flux $\sim 80$ pc from the pulsar requires rapid diffusion of high 
energy particles with 
$\tau_{esc} \approx 90 \, (R / 10 \, \rm{pc})^2 (E_e/100 \, \rm{TeV})^{-1} \, \rm{year}$,
contrary to the common assumption of toroidal magnetic fields with strong magnetic confinement.
The model predicts a rather uniform \emph{Fermi} LAT surface brightness out to $\sim1^{\circ}$ from
the pulsar, in good agreement with the recently discovered LAT source centered $0.5^{\circ}$ southwest of
PSR J1826$-$1334 with extension $0.6 \pm 0.1^{\circ}$.    

\end{abstract}

\section{Introduction}

Pulsar wind nebulae (PWNe) confine magnetized pulsar winds via a 
termination shock, whereupon particles are re-accelerated to extremely
high energies and emit radiation across the electromagnetic spectrum.  
Many PWNe are spatially resolved in the radio, X-ray, and even very
high energy (VHE) wavebands, and thereby provide an excellent laboratory to
study not only pulsar winds and dynamics, but also shock
processes, the ambient medium, magnetic field evolution, and particle transport. 

An archetypal PWN system is HESS J1825$-$137 associated with
the energetic pulsar PSR\,J1826$-$1334, which possesses an energy-dependent
TeV morphology thought to be caused by relic electrons from earlier
epochs \citep{aet06}.  This system is also a prime example
of offset VHE PWNe, with the H.E.S.S. flux centered 
$0.3^{\circ}$ south of the pulsar and VHE emission extending $1.2^{\circ}$ to the south. 
The TeV emission softens from a photon index of $\Gamma = 1.8$ near the pulsar
to $\Gamma = 2.5$ in the outer reaches of the nebula.   

PSR J1826$-$1334 was detected in the Jodrell Bank 20 cm radio survey \citep{cliftonetal92},
with period 101 ms, spin-down power 
$\dot{E} = 2.8 \times 10^{36} \, \rm{erg \, s^{-1}}$, and characteristic age
$\tau_c = 21.4$ kyr.  The dispersion measure distance of
the pulsar is $\sim 3.9$
kpc \citep{cordes+lazio02}.  
X-ray observations of the PWN with {\emph{XMM-Newton}} showed a compact core with a
hard photon index ($\Gamma = 1.6^{+0.1}_{-0.2}$) of size 30$''$ embedded
in a larger diffuse structure of extension $\sim 5'$ extending to the
south of the pulsar with a softer photon index of $\Gamma \approx
2.3$ \citep{gaensler03}.  \emph{Chandra} observations by \citet{pavlov08} revealed the compact core
to be composed of a bright inner component $\approx 7\arcsec \times 3 \arcsec$ surrounded by an outer
component elongated in the east-west direction.  More recent X-ray observations with 
\emph{Suzaku} revealed diffuse X-ray emission of $\Gamma \approx 2.0$
extending $15'$ south of the pulsar 
with little sign of spectral softening \citep{uchi09}.
HESS J1825$-$137 has not been detected in the radio \citep{osborneetal09}.

Using archival VLA observations of PSR J1826$-$1334 \citet{pavlov08} measured
a proper motion of $440 \, d_4 \, {\rm km \, s^{-1}}$ 
(with distance $d = 4 d_4$ kpc) at a position angle $\approx 100 \deg$ east of north.  
This places the pulsar birthsite some $ 8 \arcmin = 9 \, d_4 \, {\rm pc}$ to the west for an age of 20 kyr
(see Figure 1), and
rules out pulsar proper motion as the cause of the large southerly extension of HESS J1825$-$127.  
\citet{gaensler03} advocated that the 
offset is instead likely the result of an asymmetric reverse shock returning from the north, as proposed by
\citet{blondinetal01} to explain the similar structure observed in Vela-X.
Indeed, \citet{lemiereetal06} found a compelling spatial correlation between a molecular cloud discovered in $CO$
studies and the VHE source.  The distance to this cloud is measured as 4 kpc, and with a total mass of
$\approx 4 \times 10^5 \, M_{\odot}$ naturally provides the ISM inhomogeneity to the north required for the reverse
shock scenario.   The fact that emission extends roughly perpendicular to the pulsar motion suggests
that the reverse shock velocity must be significantly greater than the pulsar proper motion.  

The \emph{Fermi} Large Area Telescope (LAT) recently detected HESS J1825$-$137 
using 20~months of survey data \citep{1825fermi}.  The gamma-ray emission detected by the
LAT boasts a similar size to the H.E.S.S. source, with 
extension $0.56^{\circ} \pm 0.07^{\circ}$ for an assumed
Gaussian model. The $1-100$ GeV LAT spectrum of this source is
well described by a power-law with a spectral index of $1.4 \pm
0.2$. Single-zone spectral energy distribution (SED) modeling of the X-ray and gamma-ray data
showed that  $\approx 5 \times 10^{49}$ erg injected in the form of 
electrons evolved over $25$
kyr could reproduce the broadband aggregate spectrum, and
that the hard LAT spectrum 
was consistent with both a simple power-law
electron injection spectrum, and with a relativistic Maxwellian with a power-law tail \citep{1825fermi}.

In this paper we describe SED modeling of the X-ray and gamma-ray data, with 
particular emphasis on the spatial variations in spectra.  
In Section 2 we overview the X-ray and VHE data, Section 3  details the SED model,
Section 4 describes results of SED fitting, and Section 5 discusses the conclusions
that can be drawn from model fitting.

\section{Data Analysis}

\subsection{Suzaku Data}

Diffuse \emph{Suzaku} X-ray emission surrounding PSR J1826$-$1334 
was analyzed by \citet{uchi09}.  
The $18 \arcmin$ \emph{Suzaku} XIS field of view is broad enough to encompass the bright core of 
HESS J1825$-$137, and \citet{uchi09} reported emission nearly to the edge of all four XIS chips.  
Concurrently, a dedicated background pointing $35 \arcmin$ northeast of the initial exposure 
and outside the H.E.S.S.\ emission region was used for spectral backgrounds.  
More recently, a greater portion of the H.E.S.S.\ excess was covered 
by a grid of three \emph{Suzaku} pointings 
(unpublished at the time of submission)
to the South and West of the initial pointing (see Table 1).
A further \emph{Suzaku} exposure centered on PSR J1826$-$1334 
(also unpublished) expanded the data coverage of the PWN core.
These five images therefore cover $\approx 40 \arcmin \times  40\arcmin$ of the brightest portion of 
HESS J1825$-$137, and are listed in Table 1.  
A number of point sources are observed in the \emph{Suzaku} mosaic, five of which (A-E) 
were noted by \citet{uchi09}.  These sources are listed in Table 2, as well as displayed
in Figure 1.  Sources E and N have no cataloged counterpart.

\begin{table}[b!]
\caption{\emph{Suzaku} Data}
\begin{tabular}{c@{\hspace{0.05cm}}c@{\hspace{0.3cm}}c@{\hspace{0.3cm}}c@{\hspace{0.3cm}}c@{\hspace{0.15cm}}c@{\hspace{0.05cm}}}
\hline
\hline
No. \T \B & Obs ID & Date & Pointing\tablenotemark{a} & Location & Time (ks) \\
\hline
0 & 501045010 \T & 2006-10-19 & 276.90, -13.29 & BG & 52.1 \\
1 & 501044010 & 2006-10-17 & 276.51, -13.71 & H.E.S.S. core & 50.3 \\
2 &503028010 & 2008-10-15 & 276.50, -14.01 & S of obs \#1 & 57.2 \\
3 & 503029010 & 2008-10-17 & 276.20, -13.72 & W of obs \#1 & 57.2 \\
4 & 503030010 & 2008-10-19 & 276.19, -13.99 & SW of obs \#1 & 57.2 \\
5 & 503086010\B & 2009-03-19 & 276.57, -13.59 &  PSR & 52.1 \\
\hline
\end{tabular}
\tablenotetext{a}{ RA, Dec (degrees)}
\end{table}
\normalsize


\begin{table}[b!]
\caption{X-ray Point Sources}
\begin{tabular}{c@{\hspace{0.15cm}}c@{\hspace{0.15cm}}|c@{\hspace{0.15cm}}c}
\hline
\hline

Source \T \B & Name/Location\tablenotemark{a} & Source & Name/Location\tablenotemark{a}\\
\hline
A\T & 2XMM J182557.9-134755 & H & 2XMM J182539.9-133130 \\
B & 2XMM J182617.1-134111 & I & 2RXP J182535.2-135049 \\
C & 2XMM J182629.5-133648 & J & 2RXP J182458.3-133747 \\
D & 2XMM J182620.9-134426 & K & 2RXP J182509.9-134621 \\
E & 18:26:12 -13:48:33 & L & 2RXP J182601.4-140423 \\
F & 2RXP J182640.4-133911 & M & 2RXP J182436.5-140429 \\
G\B & 2RXP J182642.6-133625 & N & 18:24:14 -13:58:15 \\
\hline
\end{tabular}
\tablenotetext{a}{ RA, Dec (Sexagesimal)} 
\end{table}
\normalsize

We utilize the standard pipeline screened events, and analyze the XIS chips with XSelect version 2.4.  
The four most recent pointings occurred after the loss of the XIS2 chip, 
so these pointings include the two remaining front side illuminated chips (XIS0, XIS3) as well as
 the back side illuminated chip (XIS1).  All observations operate the XIS in normal clocking 
full window mode, with the data split between two editing modes: $3\times 3$ and $5\times 5$. 
An exposure corrected mosaic of the five \emph{Suzaku} exposures is shown in Figure 1.  
Diffuse emission clearly extends $\sim 18 \arcmin$
south of the pulsar position, with very little emission beyond this, or to the north of the pulsar.

\begin{figure*}[tbp!]
\epsscale{1.1}
\plottwo{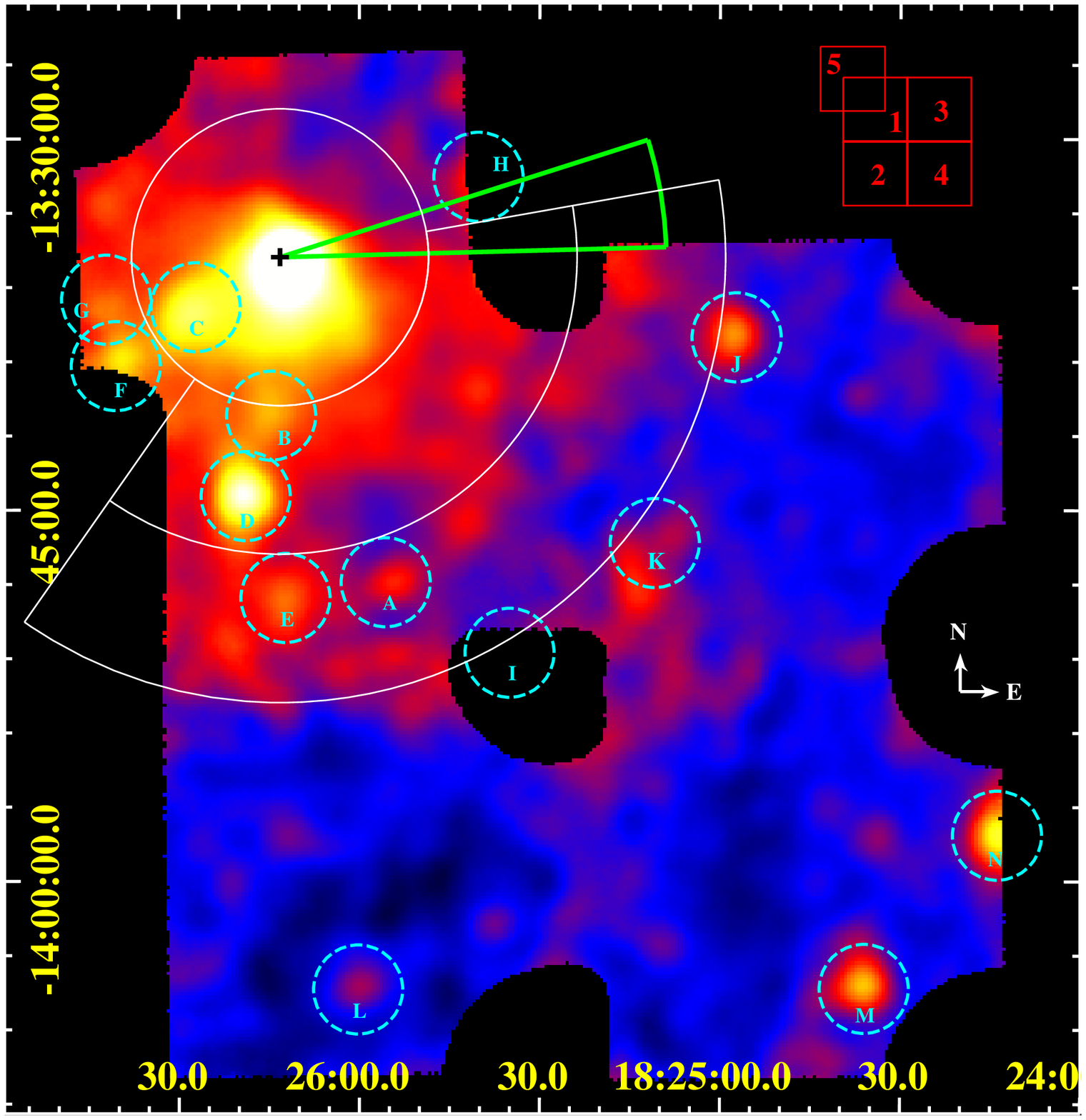}{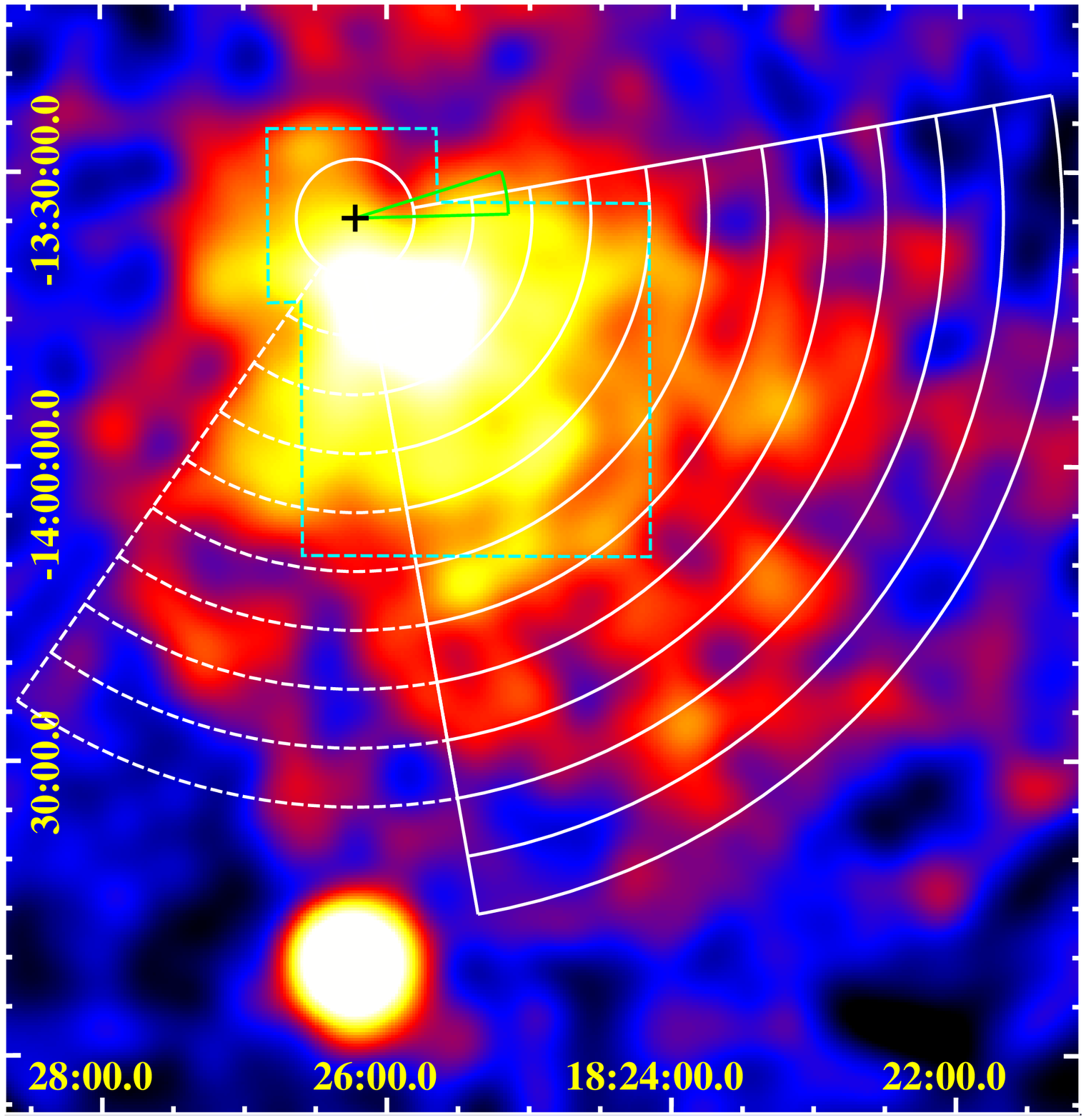}
\caption{
{\bf Left:}
\emph{Suzaku} exposure corrected $1-8$ keV counts mosaic of the front side illuminated chips for all five pointings,
smoothed with a $1.2 \arcmin$ Gaussian.  
Point sources (see Table 2) are denoted by $1.8 \arcmin$ radius dashed circles, and
the three X-ray spectral extraction regions are shown in solid white.  
PSR J1826$-$1334 is indicated by the black cross,
and the green wedge indicates the range of possible pulsar birthsites 
\citep{pavlov08}.  This green wedge
extends some $16 \arcmin$, corresponding to the presumed birthsite for an age of 40 kyr.  
The location of each of the five exposures is indicated by the map in the upper right.  
{\bf Right:}
H.E.S.S. excess counts map showing the VHE spectral extraction regions (solid white), and larger
extrapolated regions (dashed white). The \emph{Suzaku} field of
view is shown in cyan.  The bright point source to the south is the $\gamma$-ray binary LS 5039.  
}
\end{figure*}

\subsection{H.E.S.S. Data}

The twelve H.E.S.S. spectral extraction regions extend in $6 \arcmin$ chunks 
 $1.2 ^{\circ}$ from the pulsar ($0.1^{\circ} = 7.0 \, d_4 \, \rm{pc}$),
covering a $90^{\circ}$ sector from approximately $190 ^{\circ} - 280 ^{\circ}$, 
measured counterclockwise from North (\citet{aet06}, Figure 4). 
While these regions capture
the majority of the VHE emission, inspection of the H.E.S.S. excess map indicates that a sizable 
fraction of the emission
lies to the southeast of the pulsar and outside of the extraction regions. 
An accurate value of the ratio of X-ray to gamma-ray flux is important for modeling
purposes (see Section 3.1).
Therefore, to account for the missing VHE flux, 
and to allow better comparison with the \emph{Suzaku} X-ray data (which extends beyond the
$90^{\circ}$ sector) and LAT GeV gamma-ray data (which is drawn from a very large extraction region),
we extend
the quarter annuli an extra $45 ^{\circ}$ such that they now cover $145 ^{\circ} - 280 ^{\circ}$.  
Barring a full reanalysis of the H.E.S.S. data (which is not possible given
the propriety nature of H.E.S.S. data) 
the best we can do is to scale the amplitude of each slice's spectrum by
the ratio of total image counts in the original versus extended region.  
As a check of this method, we compute the image counts ratios between slices.  
For the original (unscaled) H.E.S.S. regions, the ratio of image counts between
any two regions lies within the error bounds of the published H.E.S.S. flux ratios
between those two regions.  This lends some credence to the assumption that the
flux within a region scales directly with the number of counts in the H.E.S.S.
excess map.  

The scaling factors are listed in Table 3 (along with X-ray scaling factors, 
described in Section 2.3).
These expanded regions capture 81\%
of the counts within the $0.8 ^{\circ}$ radius region used to extract the aggregate 
H.E.S.S. spectrum.
The VHE flux data points of the SED plots (Figures~\ref{fig:SED1}, \ref{fig:SED2}, 
\ref{fig:SED3}, \ref{fig:SED4})
are therefore taken to be the rescaled points from Figure 4 of
\citet{aet06}, 
and summing the rescaled flux data points 
of these regions yields 
data points $\approx 25$\%
below the aggregate H.E.S.S. data.
This is quite close to the expected $19$\% 
difference expected from the counts ratio in the H.E.S.S. excess map.  
Some of the missing flux is the northeast extension of the nebula, but some is likely 
contributed by adjacent regions to the north.  
\citet{aet06} estimate the systematic error on the flux scale to be 20\%, 
though for SED fitting we utilize statistical errors only.
Figure 1 shows the H.E.S.S. excess map and spectral extraction regions.

\begin{table}[b!]
\caption{H.E.S.S. Spectral Data}
\begin{tabular}{c@{\hspace{0.15cm}}c@{\hspace{0.15cm}}c@{\hspace{0.3cm}}c@{\hspace{0.3cm}}c}
\hline
\hline
Region \T \B & X-Ray Scaling & H.E.S.S. Scaling & $\Gamma$\tablenotemark{a} & Scaled Flux\tablenotemark{b} \\
\hline
$0 \arcmin -6 \arcmin \T$ & $1.13$ & $1.0$ & $1.83 \pm 0.09$ & $2.9 \pm 0.3$  \\
$6 \arcmin -12 \arcmin$ & $1.16$ & $1.5$ & $1.96 \pm 0.08$ & $4.4 \pm 0.4$  \\
$12 \arcmin -18 \arcmin$ & $1.27$ & $1.4$ & $2.20 \pm 0.06$ & $6.0 \pm 0.4$  \\
$18 \arcmin -24 \arcmin$ & & $1.4$ & $2.25 \pm 0.06$ & $7.4 \pm 0.4$  \\
$24 \arcmin -30 \arcmin$ & & $1.4$ & $2.32 \pm 0.07$ & $8.7 \pm 0.5$  \\
$30 \arcmin -36 \arcmin$ & & $1.4$ & $2.37 \pm 0.06$ & $9.7 \pm 0.5$  \\
$36 \arcmin -42 \arcmin$ & & $1.3$ & $2.36 \pm 0.08$ & $7.4 \pm 0.5$  \\
$42 \arcmin -48 \arcmin$ & & $1.4$ & $2.41 \pm 0.09$ & $7.6 \pm 0.6$  \\
$48 \arcmin -54 \arcmin$ & & $1.2$ & $2.42 \pm 0.09$ & $6.1 \pm 0.5$  \\
$54 \arcmin -60 \arcmin$ & & $1.2$ & $2.59 \pm 0.13$ & $6.2 \pm 0.6$  \\
$60 \arcmin -66 \arcmin$ & & $1.0$ & $2.43 \pm 0.09$ & $5.9 \pm 0.5$  \\
$66 \arcmin -72 \arcmin \B$ & & $1.0$ & $2.45 \pm 0.35$ & $2.9 \pm 0.7$  \\
\hline
\end{tabular}
\tablenotetext{a}{ Data taken from \citet{aet06}, Table 2}
\tablenotetext{b}{ $0.25-10$\,TeV fluxes in units of $10^{-12} \, {\rm erg\,cm^{-2}\,s^{-1}}$}
\end{table}

\subsection{X-ray Spectrum}

Motivated by SED modeling, we extract \emph{Suzaku} X-ray spectra from $6 \arcmin$ wedges to mimic the 
regions of \citet{aet06}. Spectral fitting is accomplished with Xspec version 12.6.
For background subtraction we follow the same procedure as \citet{uchi09} 
(a more in-depth discussion can be found in this paper) and 
define the same region as these authors: a $9 \arcmin$ circle on 
the background exposure, 
excising $1.8 \arcmin$ radius circles around the four point sources noted by these authors. 
The non-uniformity of the background is taken into account internally within Xspec.  
Spectra are extracted with XSelect, while ARF and RMF response files  
are generated with the
xisrmfgen and xissimarfgen functions, which internally 
take into account the different responses of each chip.
The ASCA FTOOL addascaspec is utilized to combine 
the spectra and responses of the XIS front side illuminated chips.  
The XIS1 back side illuminated chip possesses a markedly different response function from the 
front side illuminated chips, and so is analyzed in parallel rather than added to the other chips.  

Initially, we extract and fit the spectrum of the $1.5 \arcmin$ circle surrounding PSR J1826$-$1334 
titled Region A by 
\citet{uchi09}.  Fitting an absorbed power-law model we achieve a fit well within errors of the 
parameters reported by these authors
($\Gamma = 1.7 \pm 0.1$, unabsorbed $1-8$ keV flux of 
$1.2 \pm 0.2 \times 10^{-12} \, \rm{erg \, cm^{-2} \, s^{-1}}$).
We subsequently define regions mimicking the inner $6 \arcmin$ circle 
surrounded by $6 \arcmin$ thick  partial annuli shown in Figure 1.  From these regions we excise 
$1.8 \arcmin$ radius circles around the 5 point sources defined by \citet{uchi09}. 
For the innermost $6 \arcmin$ circle 
we use both the original dataset (501044010), as well as the most recent dataset (503086010)
and jointly fit the absorption
and power-law index.  Since the original dataset does not encompass the entirety of the $6 \arcmin$ circle 
we allow the power-law normalization to vary
between the two datasets, though we quote the flux from the entire circle.  
The $6 \arcmin - 12 \arcmin$ region follows the same procedure, using both datasets 501044010
and 503086010.
For overlapping regions (exposure 1 and exposure 5) the joint spectral fits to
both exposures are completely
consistent (albeit with smaller errors) in both index and total flux (when taking
into account the differing areas of the regions) with fits to the individual exposures.  

All fits are to an absorbed power-law from $1-8$ keV with quoted single parameter 68\%
($1\sigma$) errors.  
We fit the $6 \arcmin - 12 \arcmin$ region
with $N_H$ fixed at the value determined in the innermost region, and find values
completely consistent (albeit with smaller errors) with fitting with $N_H$ free.  
Though $N_H$ may vary over the nebula, we see no evidence of this, and IR maps
show no appreciable gradient over the nebula, so for simplicity we adopt a uniform
column density.
The pulsar will contribute flux to the innermost
region, though with a $1-8$ keV flux of
$\sim 2 \times 10^{-14} \, \rm{erg \, cm^{-2} \, s^{-1}}$
\citep{pavlov08}, the pulsar presents only a minor perturbation to the PWN flux.  

The $12 \arcmin - 18 \arcmin$ wedge 
stretches across three different pointings
and suffers from a much lower surface brightness than the inner two regions, greatly
complicating flux extraction from this region.  
Due to the limited statistics, we fix the absorption to the best fit value from the innermost region.  
Extracting flux from the portion of this region which lies within the original
pointing  (501044010) yields a $\sim 5 \sigma$ detection.  
No significant emission further south and west of the pulsar is seen in any
of the three exposures.  
In fact, we can use the apparent lack of flux in the southwest exposure furthest from the pulsar (503030010) 
to estimate the uncertainty in the background.  Selecting a background region from this chip alters
flux estimates by $\approx 1$\% for Region 1 ($0 \arcmin - 6 \arcmin$),
 $\approx 3$\% for Region 2, and $\approx 26$\% for Region 3.  
An additional source of error stems
 from the uncertainty in the ARF calculation for complex extended sources, which leads to uncertainty
 in the source flux \citep{ishisakietal07}, which we estimate at $\approx 10$\%.  
Note that background subtraction errors dominate for Zone 3; in fact, the relatively hard spectrum
fit here may reflect residual background contamination.  
In the SED plots we show regrouped X-ray data with the ARF and background systematic errors
summed in quadrature with $1\sigma$ statistical errors.

To be perfectly consistent, for modeling purposes we should utilize identical
regions for both X-ray and VHE gamma-ray data.  The excised point sources and 
slight portion of the H.E.S.S. spectral extraction regions which extends beyond
the \emph{Suzaku} exposures therefore require a small adjustment to the X-ray
flux level.  The raw X-ray data is therefore rescaled by the fraction of missing
area, with the scaling factors shown in Table 3.  
Table 4 and the SED plots therefore display the rescaled fluxes (with errors
adjusted accordingly) in order to remain fully consistent with the H.E.S.S. data.

\begin{table}[tb!]
\caption{X-Ray Spectral Fits}
\begin{tabular}{c@{\hspace{0.4cm}}c@{\hspace{0.4cm}}c@{\hspace{0.4cm}}c@{\hspace{0.4cm}}c}
\hline
\hline
Region \T \B & $N_{H}$\tablenotemark{a} & $\Gamma$ & Unabs. Flux\tablenotemark{b} & $\chi^2$/d.o.f. \\
\hline
$0 \arcmin -6 \arcmin \T \B$ & $1.26 \pm 0.06$ & $1.93 \pm 0.05$ & $4.90_{-0.32}^{+0.34}$ & 196/249 \\
\hline
$6 \arcmin -12 \arcmin \T$ & $1.12_{-0.10}^{+0.11}$ & $2.03 \pm 0.09$ &  $3.05_{-0.34}^{+0.39}$ & 144/173 \\
$6 \arcmin -12 \arcmin$ \B & $1.3$\tablenotemark{c} & $2.17 \pm 0.05$ & $3.18 \pm 0.15$ & 146/174 \\
\hline
$12 \arcmin -18 \arcmin \T \B$ & $1.3$\tablenotemark{c} & $1.78 \pm 0.18$ & $1.04_{-0.19}^{+0.22}$ & 44/59 \\
\hline
\end{tabular}
\tablenotetext{a}{ interstellar absorption $\times10^{22} \, \rm{cm}^{-2}$}
\tablenotetext{b}{ $1-8$\,keV fluxes in units of $10^{-12} \, {\rm erg\,cm^{-2}\,s^{-1}}$}
\tablenotetext{c}{ held fixed} 
\label{pwnspec}
\end{table}

The X-ray flux falls off rapidly with radial distance from the pulsar, likely the result
of both lower magnetic fields as well as cooling losses, as discussed below.  
Figure 2 shows a radial profile of fluxes in the X-ray and TeV energy bands, clearly indicating the 
much greater extent of the nebula in the H.E.S.S. regime.  For this figure, we 
fix $N_H$ at $1.3\times 10^{22} \, \rm{cm^{-2}}$ and extract X-ray spectra
from $3\arcmin$ width regions for greater detail than the $6\arcmin$ regions shown in Table 4.

\begin{figure}[tpb!]
\begin{center}
\end{center}
\epsscale{1.2}
\plotone{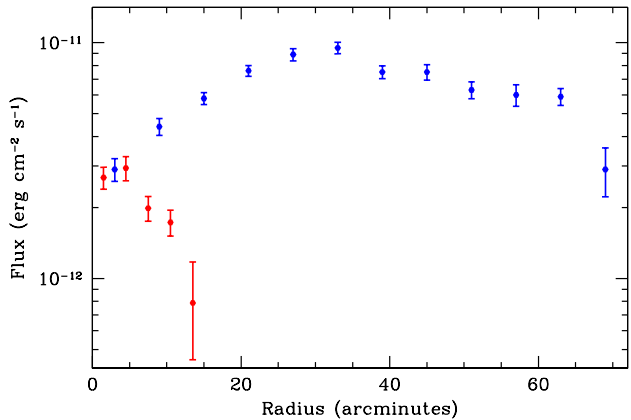}
\caption{
Radial flux profile with $1\sigma$ statistical error bars.  
\emph{Suzaku} $1-8$ keV X-ray flux is shown in red, while
the far greater extent of the H.E.S.S. 0.25 - 10 TeV flux (from Table 3) is in blue.
}
\end{figure}

\newpage

\section{SED Modeling}

\subsection{Motivation}

\citet{1825fermi} applied a one-zone SED model to the new \emph{Fermi} LAT data of HESS J1825$-$137 and found an adequate fit to
the aggregate gamma-ray and X-ray data.  One-zone models can help constrain certain global nebula properties, such
as the total energy injected in the form of electrons, the electron spectrum in the uncooled regime, and the mean 
magnetic field.  Yet such models by definition assume a completely homogeneous nebula, 
contrary to the differing X-ray and VHE sizes  
and spectral steepening observed in both X-rays and gamma-rays.
To further investigate these phenomena we implement a spatially resolved model of
HESS J1825$-$137.

We begin by investigating how simple scaling relations inform model implementation.
Starting assumptions include: the source of all particles and magnetic field in 
the nebula is PSR J1826$-$1334, ambient photon fields are static, and the nebula
is 4 kpc distant.
Let us first consider the electrons present in the innermost zone (Zone 1) adjacent the pulsar.
In a transverse magnetic field these particles synchrotron 
radiate photons at mean energy: 
\begin{eqnarray}
E_{\gamma} \approx 2.2 (E_e/100 \, {\rm TeV})^2 (B/10 \, \mu {\rm G}) \, \, {\rm keV.}
\end{eqnarray}
We scale with a low field of $10 \, \mu {\rm G}$ since, as we shall see later, even in the innermost
zone the mean field is only $\sim 10 \, \mu {\rm G}$ for the best fit model.
Such particles cool rapidly, with a lifetime ($\tau \equiv E/\dot{E}$) of:
\begin{eqnarray}
\tau_{sync} \approx 820 \, (E_e/100 \, {\rm TeV})^{-1} (B/10 \, \mu {\rm G})^{-2} \, \, {\rm year.}
\end{eqnarray}
Therefore the cooling timescale for electrons radiating synchrotron photons at 
mean energy $E_{\gamma}$ is:
\begin{eqnarray}
\tau_{sync} \approx 1200 \, (E_{\gamma}/1 \, {\rm keV})^{-1/2} (B/10 \, \mu {\rm G})^{-3/2} \, \, {\rm year.}
\end{eqnarray} 
The flat ($\Gamma \approx 2$) X-ray spectrum extending to 10 keV of Region 1 indicates that the electrons 
responsible for this emission are largely uncooled, implying 
$\tau \lesssim \tau_{sync} \approx 380 \, (B/10 \, \mu {\rm G})^{-3/2} \, \, {\rm year}$.
The inner region therefore must be dominated by particles recently injected by
the pulsar.  
In the uncooled regime the synchrotron spectral flux scales with the magnetic field and
number of electrons as: $F(E)_{sync} \propto n_{e}\times B^{3/2}$.  Over timescales
short compared to the 21 kyr spin-down age of the pulsar the  
number of electrons $n_e$ in Zone 1 
is roughly proportional to the time period over which electrons have
been injected $\delta t$, so $F(E)_{sync} \propto \delta t \times B^{3/2}$, a relation
we will return to momentarily.

Electrons inverse Compton (IC) upscatter CMB photons to a mean energy of: 
\begin{eqnarray}
E_{\gamma} \approx 0.32 \, (E_e/10 \, {\rm TeV})^{2}  \, \, {\rm TeV.}
\end{eqnarray}
For young electrons IC cooling is of little consequence, since 10 TeV electrons scattering 
off the CMB
cool in $\tau_{IC} \approx 140 \, (E_e/10 \, {\rm TeV})^{-1} \, \, {\rm kyr}$.
We scale to 10 TeV rather than 100 TeV since at 100 TeV the Thomson regime is no longer realized and 
 Klein-Nishina corrections must be incorporated.  For example, while the above scaling relation for
inverse Compton cooling holds for electrons with $E_e \leq 10$ TeV, 100 TeV electrons cool 
slowly over $36$ kyr,
rather than the $14$ kyr predicted by this relation.   
We therefore consider only synchrotron cooling in the following discussion.
The lowest energy H.E.S.S. point 
at $E_{\gamma} \approx 0.3$ TeV requires electrons
of energy $E_e \approx 10$ TeV,
which synchrotron cool slowly at a rate of
$\tau_{sync} \approx 8200 \, (B/10 \, \mu {\rm G})^{-2} \, \, {\rm year}$.
Therefore, electrons responsible for the low energy H.E.S.S. points
cool quite slowly compared to those responsible for the X-ray flux.  

The electrons responsible for the lowest energy H.E.S.S. points (and therefore  
all the \emph{Fermi} LAT flux points as well) are radiatively uncooled over time spans of a few
thousand years, and provide a good indication of just how many electrons are present
in the inner nebula.  This is due to the fact that for $E_e \leq 10$ TeV the Thomson limit holds and
IC flux scales linearly with the number of electrons ($n_e$) present: $F(E)_{IC} \propto n_e$.  
The exact number of electrons at 10 TeV 
will depend on the electron injection spectrum parameters as well as the injection time $\delta t$, and
very weakly on the magnetic field due to feeble synchrotron cooling, so
$F(E)_{IC} \propto \delta t$.    
The X-ray data breaks the degeneracy in magnetic field since $F(E)_{sync} \propto \delta t \times B^{3/2}$, 
so between X-ray and gamma-ray data the electron injection spectrum, age, and
magnetic field are very well constrained.  For modeling the data, this means
that once an injection spectrum is selected, the gamma-ray flux determines the injection time, while
the ratio of X-ray to gamma-ray
flux determines the magnetic field.   
 
\subsection{Pulsar Spin-down}
 
As pulsars spin down, they dissipate rotational kinetic energy according to:
\begin{eqnarray}
\dot{E} = I \Omega \dot{\Omega},
\end{eqnarray} 
with $\Omega$ the angular frequency and $I$ the neutron star's moment of inertia, assumed to be $10^{45} \, \rm{g \, cm^2}$.  
This energy goes into a magnetized particle wind such that for a braking index of $n$:
\begin{eqnarray}
\dot{\Omega} \propto \Omega^n .
\end{eqnarray}
For spin-down via magnetic dipole radiation $n=3$, though $n$ has only been confidently measured
for five pulsars, in each case falling between $2 < n < 3$ (\citet{livingstoneetal07} and references therein).
Integrating Equation 6 yields the age of the system \citep{1977puls.book}:
\begin{eqnarray}
T = \frac{P}{(n-1) \mid \dot{P} \mid} \left( 1- \left( \frac{P_0}{P} \right)^{(n-1)} \right) ,
\end{eqnarray}
where $P_0$ is the initial spin period, and $\dot{P}$ the period derivative.
For $P_0 \ll P$ and $n=3$ this equation reduces to the characteristic age of the pulsar $\tau_c \equiv P/2\dot{P}$.
The spin-down luminosity of the pulsar evolves as \citep{pc73}:
\begin{eqnarray}
\dot{E}(t) = \dot{E_0} \left( 1 + \frac{t}{\tau_{0}} \right) ^{-\frac{(n+1)}{(n-1)}},
\end{eqnarray}
with the initial spin-down timescale defined as:
\begin{eqnarray}
\tau_{0} \equiv \frac{P_0}{(n-1) \mid \dot{P_0}  \mid} \, .
\end{eqnarray} 
Given that the current $P$, $\dot{P}$, and $\dot{E}$ are known, once an initial period $P_0$ and braking
index $n$ are selected 
the age $T$ and spin-down history of the system are determined according to the equations above. 

The energy flux in the magnetized wind $\dot{E}$ is split between 
electromagnetic $\dot{E_B}$ and particle energy $\dot{E_{e}}$, with:
\begin{eqnarray}
\dot{E}_{e} & = & \eta \dot{E} \nonumber \\
\dot{E}_{B} & = & (1 - \eta) \dot{E} .
\end{eqnarray}
Their ratio is commonly referred to as the wind magnetization parameter: 
\begin{eqnarray}
\sigma \equiv \dot{E}_B / \dot{E}_{e} = (1-\eta)/\eta ,
\end{eqnarray}
with $\sigma$ typically thought to be $\ll 1$ past the termination shock.
The particle energy is presumed to go into an electron/positron plasma with an exponentially 
cutoff power-law spectrum:
\begin{eqnarray}
\frac{dN}{dE} \propto E^{-p} e^{-E/E_{cut}} \; \; \; \; \; E > E_{min} ,
\end{eqnarray}
with $E_{min}$ the minimum particle energy.  We initially select $E_{min}$ as 100 GeV, which 
is well within the realm of minimum particle energies considered by \citet{kc84b} in 
magnetohydrodynamical modeling of the Crab Nebula.
The normalization of this power-law varies with time following
Equation 8, though we treat the index, minimum energy, and cutoff energy as static. 
We begin by modeling pure power-law injection, although we later consider an injection spectrum
composed of a relativistic Maxwellian with a non-thermal tail, as proposed by \citet{spitkovsky08}.

\subsection{Spatial Evolution}

Pulsar winds flow out at relativistic speeds until the ram pressure of the wind 
is balanced by the internal pressure of the PWN at the termination shock.
High-resolution X-ray images can help illuminate the location of this termination shock, and
 Chandra images of the pulsar reveal that the brightest inner PWN component
stretches some $7 \arcsec \times 3 \arcsec$ ($0.14 \times 0.06 \, \rm{pc^2}$ at 4 kpc) \citep{pavlov08},
with the short axis roughly oriented toward the pulsar birthsite.  The overall extent of the 
equatorial flow is generally $2-3$ times larger than the termination shock radius \citep{ng+romani04},
and we accordingly adopt a value of $r_{ts} = 0.03$ pc for the radius of the termination shock.
 
Past the termination shock
the pulsar wind expands radially at a much reduced velocity.  
In the model of of \citet{kc84a} 
the rapid rotation of the pulsar wraps up magnetic field lines, resulting
in a primarily toroidal field at and beyond the termination shock.  The field 
direction is therefore approximately perpendicular to the wind flow velocity.  
Since particles effectively cannot
cross toroidal magnetic field lines, the bulk flow of particles predicts 
an ordered layering of particle ages with increasing distance from the pulsar, of age:
\begin{eqnarray}
t(r) = \int_0^r v(r')^{-1} \, dr' .
\end{eqnarray}
with $v(r)$ the velocity profile.  

On larger scales, we consider the evolution of the parent supernova remnant (SNR) of PSR J1826$-$1334.
Early in the pulsar lifetime the SNR expands freely into the ambient medium, typically
at a velocity of $\sim 10,000 \, \rm{km \, s^{-1}}$.
The large $\sim 80$ pc size of the VHE PWN and lack of any observed SNR shell 
led \citet{dejager+datai08} to propose a very large SNR of radius  
$\approx 120$ pc resulting from an energetic supernova explosion 
($E_{SN} = 3 \times 10^{51}$ erg) of age $\approx 40$ kyr,
 and a very low ambient medium density of 
$n_0 \approx 0.001 \, \rm{cm^{-3}}$. 
Early on, a spherically symmetric PWN expands rapidly into unshocked supernova ejecta 
with nearly constant velocity
\citep{vander01}:
\begin{eqnarray}
v_{pwn} & \approx & 3100  \left( \frac{\dot{E_0}} {10^{39} \, \rm{erg \, s^{-1}}} \right)^{1/5} 
	\left( \frac{E_{SN}} {3 \times 10^{51} \, \rm{erg}} \right)^{3/10} \nonumber \\
	& & \times
	\left( \frac{M_{ej}} {8 M_{\odot}} \right)^{-1/2} 
	\left( \frac{t} {1 \, \rm{kyr}} \right)^{1/5} \; \rm{km \, s^{-1} .}
\end{eqnarray}
As the SNR expands and sweeps up more mass, it eventually
reaches a point where the swept up ISM material is 
approximately equal in mass to the mass of the supernova ejecta.  
At this point the SNR enters the Sedov-Taylor phase as the swept-up
material begins to dominate the SNR dynamics.  
Adopting a uniform density ISM, for the standard 
core-collapse supernovae 
assumption of a constant density inner core surrounded by a 
$\rho \propto r^{-9}$ outer envelope the transition to 
the Sedov-Taylor phase occurs at \citep{truelove&mckee99}:
\begin{eqnarray}
t_{S-T} & \approx & 7100 \, \left( \frac{E_{SN}}{3 \times 10^{51} \, \rm{erg}} \right) ^{-1/2}  
	\left( \frac{M_{ej}}{8 M_{\odot}} \right)^{5/6} \nonumber \\
	& & \times
	 \left( \frac{n_0}{0.001 \, \rm{cm^{-3}}} \right) ^{-1/3}  \, \rm{year.}
\end{eqnarray}
In this phase a reverse shock is driven deep into the SNR interior, 
and for a SNR expanding
into a homogenous medium the reverse shock will 
impact the PWN, crushing it and leading to a series of reverberations
which can last many thousands of years, as
demonstrated compellingly in \citet{gelfandetal09}. Calculations by
\citet{vander04} yielded
an upper limit on the reverse shock collision with
the PWN at $t_{rs} < 5 \, t_{S-T}$.
A supernova explosion into an evacuated cavity of $n_0 = 0.001 \, \rm{cm^{-3}}$ therefore
yields a maximum PWN-reverse shock collision time of
$\approx 35$ kyr (nearly double the characteristic age of the pulsar),
implying that the PWN could to this day be rapidly expanding to the southwest
 into unshocked SNR ejecta.
Of course, reflection off of the molecular cloud discovered by
\citet{lemiereetal06} could result in a reverse shock collision
on a timescale reduced by an order of magnitude from this estimate.

Axisymmetric 2-D simulations by
\citet{blondinetal01} showed that SNR expansion into an inhomogeneous medium leads
to a PWN offset from the pulsar position and mixing of the thermal
gas of the supernova ejecta and the relativistic gas of the PWN.
Rayleigh-Taylor instabilities
aid in the mixing and disruption of the nebula as bubbles of swept up material
penetrate the PWN. 
The simulations of \citet{blondinetal01} showed a cometary nebula 
$\sim 20$ kyr after the reverse shock interaction, and
nearly complete disruption of the nebula after
$\sim 50$ kyr.  These simulations do not take into account the
magnetic field that is mixed in with the relativistic particles, and which
can dampen instabilities along magnetic field lines.
Indeed, recent simulations by \citet{bucc04} suggest that the 
growth of instabilities is highly suppressed for high $\sigma$ flows,
and significant structure may exist even some $\sim 20$ kyr
after reverse shock interaction. This suggests that the ordered
layering of particle ages predicted by Equation 13
may remain valid in the case of an anisotropic reverse shock.

If one invokes PWN expansion within a freely expanding
SNR to account for the $\sim 80$ pc size of the PWN, 
a mean velocity of $3100 {\rm \, km \, s^{-1}}$ over 25 kyr is required.  
While this velocity is quite high for a PWN, it is permitted by Equation 14 for a
very low density ambient medium.
The $17 \arcmin \, (20 d_4 \, \rm{pc})$ displacement of the 
H.E.S.S. centroid south of PSR J1826$-$1334 \citep{aet06}
cannot be explained by pure expansion, however, and requires some mechanism to achieve this 
offset.  

One possibility for the southerly 
displacement is the existence of a powerful jet rapidly transporting particles far from the pulsar.
X-ray images show
no indication of any such feature, however.  In addition, PWN jets are typically aligned
with the pulsar spin axis, and a southern jet would be roughly perpendicular to the 
inferred pulsar proper motion, also typically aligned with the pulsar spin axis.   

A more palatable alternative is that the southerly offset
from the pulsar is due to crushing by the reverse shock. 
In this picture the reverse shock quickly reflects off
of the molecular cloud to the north, and then impacts
the PWN in a few thousand years.
The mechanism
responsible for the PWN displacement, be it direct interaction with the
reverse shock or a resultant anisotropy from the shock,
is evidently still at work since the young X-ray emitting electrons are 
clearly concentrated south of the pulsar.  
For a reverse shock returning from the North, particles initially
north of the pulsar will be displaced south and mixed throughout the nebula.
The fact that the H.E.S.S. spectral index of HESS J1825$-$137
increases monotonically from the pulsar 
hints that such mixing is likely far from complete.  
Precisely how the reverse shock crushing and displacement alters any ordered layering 
of the PWN requires detailed 2-D or 3-D magnetohydrodynamical simulations and 
is beyond the scope of this paper.
We ignore such effects, and hence tacitly assume
 that for HESS J1825$-$137 reverse shock interactions serve primarily to 
displace rather than compress and mix the nebula.

\subsection{Morphology}

The shape of the nebula as viewed from earth resembles a circular sector.  
While the 3-dimensional morphology of HESS J1825$-$137 is uncertain, if the morphology results from
a shock returning from the north, this implies nebular 
axial symmetry about the normal to the shock front.  One simple geometry 
which matches the \emph{Suzaku} and H.E.S.S. morphology is a spherical wedge of opening
angle $135^{\circ}$ (solid angle 3.88 steradians)
with symmetry axis along the H.E.S.S. symmetry axis $17^{\circ}$ counterclockwise
from North.  The molecular cloud which
presumably gives rise to the assymetric reverse shock is located at the same 4 kpc distance
as the pulsar, 
so we assume that the shock normal is perpendicular to the line of sight.  
For modeling purposes we split this wedge into twelve 3-D zones which mimic the spectral extraction regions.
Given the axisymmetry, the twelve spatial zones
form a series of nested bowls, or partial spherical wedges, of thickness $7.0 d_4$ pc. 
Similar to the spectral extraction regions,
we label Zone 1 adjacent to the pulsar (extending $0-7$ pc),
 and Zone 12 the outermost spherical wedge ($77-84$ pc from the pulsar).
Henceforth we refer to Regions as the 2-D projections on the sky of the 3-D spatial Zones.
Figure 3 demonstrates the adopted morphology, with the symmetry axis oriented vertically
for clarity.  
The final volume of each bubble is determined by the spherical wedge morphology discussed above.  
These volumes are listed in Table 5.
\begin{figure}[tbp!]
\begin{center}
\end{center}
\epsscale{1.}
\plotone{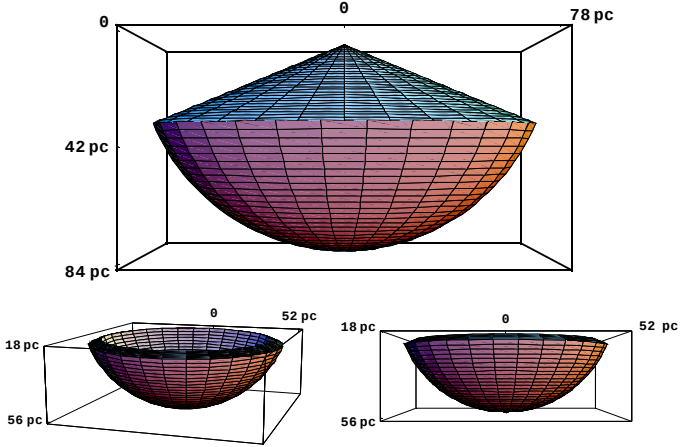}
\caption{
{\bf Top:} Spatial model of HESS J1825$-$137 as seen from Earth, with the symmetry axis
oriented vertically.  The pulsar is located at the origin, and the 
axes indicate the extent of nebula in parsecs.  
{\bf Bottom Left:} One of the twelve spatial zones (Zone 8: 49 - 56 pc from the pulsar)
plotted with a slightly elevated viewpoint to demonstrate the concave nature of the zone.
{\bf Bottom Right:} The same spatial zone seen edge-on from the Earth line of sight.
}
\end{figure}

When projected into the sky, the counts from each bowl will overlap into multiple spectral extraction regions.
In order to determining the degree of overlap we perform a Monte Carlo integration by projecting the 3-D spatial 
zones onto the plane of the sky.  
For example, consider Zone 8 extending 
$49 \, \rm{pc} - 56 \, \rm{pc}$ from the pulsar.  
We populate this zone with a large number of random points, then project each point onto
the appropriate plane defined by the viewing angle.  We then overlay the spectral extraction regions
onto this projection and determine the fraction of counts which lie within each bin. 
Assuming each bin is spatially uniform, the fraction of counts within each region is equivalent to the fraction of flux.
For all zones save the innermost, significant flux from each spatial zone is found in the interior
spectral extraction regions, as demonstrated by Figure 4. 
Table 5 shows the fraction of flux from each zone which falls
within each spectral region. 

\begin{figure}[tpb!]
\begin{center}
\end{center}
\epsscale{0.9}
\plotone{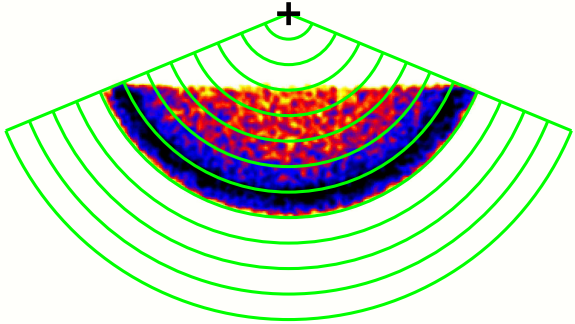}
\caption{
Smoothed projected counts from Zone 8 extending 
$49 - 56$ pc from the pulsar.
The $6'$ width spectral extraction regions are overlaid, demonstrating the overlap of counts
on the interior regions. The location of the pulsar is marked with a cross.
}
\end{figure}

\begin{table*}[tb!]
\begin{center}
\caption{Projection of Spatial Zones onto Spectral Extraction Regions}
\centering
\begin{tabularx}{1.0\textwidth}{lZZZZZZZZZZZZc}
\hline\hline
\T \B & Reg 1 & Reg 2 & Reg 3 & Reg 4 & Reg 5 & Reg 6 & Reg 7 & Reg 8 & Reg 9 & Reg 10 & Reg 11 & Reg 12 & V (pc$^3$) \\
\hline
Zone 1 $\T$ &      1.00 & & & & & & & & & & & & 440    \\
Zone 2 &      0.14 &      0.86 &  & & & & & & & & &  & 3100  \\
Zone 3 &      0.01 &      0.30 &      0.69 &  & & & & & & & &  & 4800  \\
Zone 4 &           &      0.07 &      0.34 &      0.59 &  & & & & & & & &  16000  \\
Zone 5 &           &      0.01 &      0.13 &      0.34 &      0.52 &  & & & & & & &  27000  \\
Zone 6 &           &           &      0.05 &      0.15 &      0.32 &      0.47 &  & & & & & &  40000  \\
Zone 7 &           &           &      0.02 &      0.08 &      0.16 &      0.30 &      0.44 &  & & & & & 56000 \\
Zone 8 &           &           &          &      0.04 &      0.09 &      0.16 &      0.30 &      0.41 &  & & & &  75000  \\
Zone 9 &           &           &           &      0.02 &      0.06 &      0.10 &      0.15 &      0.28 &      0.39 &  & & &  96000  \\
Zone 10 &          &           &           &      0.01 &      0.04 &      0.07 &      0.10 &      0.15 &      0.27 &      0.37 &  & &   120000   \\
Zone 11 &          &           &           &           &      0.02 &      0.05 &      0.07 &      0.10 &      0.15 &      0.26 &      0.35   & &  150000 \\
Zone 12 \B &          &           &           &           &   0.01 &      0.03 &      0.05 &      0.07 &      0.10 &      0.15 &      0.25   &   0.34 & 180000 \\
\hline
\end{tabularx}
\tablecomments
{
\centering
\scriptsize
Each row gives the fraction of flux from each spatial zone falling within each spectral region.  
For example, Column 1, Row 2 indicates the fraction of flux from Zone 2 
which falls within Region 1 (0.141). The flux within Region N is the sum of the fluxes in 
zones 1-12 times the coefficients ($C_{i,N}$) in column N, $F_N = \sum_{i=1}^{12} C_{i,N} \,F_i  $.
}
\end{center}
\end{table*}
\normalsize

\subsection{Spatial Model}

Motivated by the morphology adopted in the previous section,
we treat the PWN as a series of twelve expanding bubbles
with the radial extent of each bubble 
accounted for by a preferential southerly expansion.  
As per the discussion in Section 3.3, this southerly expansion is assumed to result from
an asymmetric reverse shock returning from the north.  
Particles are injected into each bubble for a period of time $\delta t_i$, after which the bubble
is assumed to detach from the pulsar and direct
injection of particles is discontinued.
Given the distinct possibility that in the southerly direction SNR is still freely expanding 
(leading to nearly free expansion for the PWN), for simplicity
 we assume constant 
velocity expansion of the outer boundary of the nebula $v_{outer} = 84 \, \rm{pc} / T$, 
with $T$ the pulsar age determined by Equation 7.  
Despite the adoption of a constant outer expansion velocity, particles will still
decelerate over time as they traverse the nebula and encounter supernova ejecta.  
In the interior of the nebula we therefore consider a radial velocity profile
\begin{eqnarray}
v(r,t) = v_{outer} \, \left( \frac{r}{R_{outer}} \right) ^{\alpha} \, \left( \frac{t}{T} \right) ^{-\alpha} ,
\end{eqnarray}
with $\alpha$ a fit parameter, and the time dependence 
determined by the requirement that $v_{outer}$ remain constant.
Integrating Equation 16 yields the position of various bubbles with time.
The final extent of each bubble (
e.g. Zone 1 covers $0-7$ pc from the pulsar) determines the required injection time for each bubble
via Equation 13.  
  
Once injection ends, the bulk 
evolution of each bubble is primarily adiabatic, as
 IC cooling is relatively weak and the magnetic field is reduced to the 
point that synchrotron losses are severe only for the highest energy particles.
For relativistic particles with adiabatic index $4/3$
the particle pressure is $P_{pwn} = E_e/(3 V_{pwn})$, and for purely adiabatic expansion
$P_{pwn} V_{pwn}^{4/3}$ = constant.  For all time steps particle
adiabatic losses are therefore computed via 
\begin{eqnarray}
E_e \propto V_{pwn}^{-1/3} .
\end{eqnarray}
The sound speed in the relativistic gas of the PWN ($c/\sqrt{3}$) is far greater than the 
expansion velocity of the PWN, implying that pressure differences will rapidly equilibrate and
that the PWN is nearly isobaric.  
The energy lost due to adiabatic expansion goes into accelerating supernova ejecta at the 
PWN-SNR interface, and so we consider adiabatic losses to be shared equally among all
particles present in the nebula, with the fraction of energy lost by each particle a function only 
of the volume evolution of the aggregate nebula.

\subsection{Magnetic Field Model}

In principle, one can use the injected magnetic field energy (Equation 10)
 and spatial evolution of the PWN zones
to compute the magnetic field ($E_B = \frac{B^2}{8 \pi} V$, see \citet{gelfandetal09}).
Due to the uncertainty in how the magnetic field is distributed and transported, and
the possibility of field amplification in the post-shock flow, we 
 adopt a somewhat simpler magnetic field evolution.   
 
A static and uniform magnetic field presents the simplest possible magnetic field structure,
yet the nebular magnetic field is expected to decrease from the termination shock to the far reaches 
of the nebula, as well as decrease with time as the pulsar spins down.   
Near the pulsar one common simple formulation is a dipolar magnetic field ($B \propto r^{-3}$) 
out to the pulsar light cylinder, with a toroidal field ($B \propto r^{-1}$) out to the termination
shock.  
The surface magnetic field is estimated from the period and period derivative as
\begin{eqnarray}
B_s = 3.2 \times 10^{19} \, (P \dot{P})^{1/2} \, \rm{G} .
\end{eqnarray}
For PSR J1826$-$1334 this gives a current value of $B_s = 2.8 \times 10^{12} \, \rm{G}$ for
a surface radius $r_s = 10$ km.  The light cylinder is currently located at
$r_{lc} = c/\Omega = 4.8 \times 10^8 \, \rm{cm}$. At the wind termination shock the field is 
amplified by a factor of $\sim 3$, which gives $B_{ts} \approx 400 \, \mu \rm{G}$ for 
$r_{ts} = 0.03$ pc, and the assumed geometry of $B \propto r^{-3}$ out to the light cylinder
and $B \propto r^{-1}$ from the light cylinder to the termination shock.  

The termination shock occurs at radius:
\begin{eqnarray}
r_{ts} = \left( \frac{\dot{E} \epsilon}{4 \pi c P_{pwn}}  \right) ^{\frac{1}{2}} ,
\end{eqnarray}
where $\epsilon$ is the filling factor of the pulsar wind.  For an isotropic
wind $\epsilon = 1$, though more common is a scaling with co-latitude 
$\theta$ of $\rm{sin}^2 \theta$, yielding $\epsilon = 3/2$ \citep{bog+khan02}.
The key feature here is that 
the termination shock radius scales as $r_{ts} \propto (\dot{E}/P_{pwn})^{1/2}$.  
One can estimate the PWN pressure 
as $P_{pwn} \propto \rho_{ej} v^2$ (Equation 26), 
with $\rho_{ej} \propto t^{-3}$ (Equation 24). 
The assumed constant velocity expansion therefore gives $P_{pwn} \propto t^{-3}$,
which is identical to the $P_{pwn} \propto t^{-3.0 }$ scaling computed by the \citet{gelfandetal09}
model prior to reverse shock interactions.
The termination shock
radius accordingly scales roughly as  $r_{ts} \propto \dot{E}(t)^{1/2} t^{3/2}$, and for a braking index 
near $n=2$ (where $\dot{E}(t) \propto t^{-3}$) the termination shock remains approximately static.  
Furthermore, since our SED model focuses on emission from the entire nebula
(which dwarfs $r_{ts}$), the precise value of $r_{ts}$ has little effect on the resultant SED.  For simplicity,
we therefore assume a constant value of $r_{ts} = 0.03$ pc.  

Adopting a dipolar followed by toroidal magnetic field geometry out to the termination shock, 
$B_{ts} = 3 B_{lc} \, r_{lc} \, r_{ts}^{-1} =
 3 B_s \, r_s^3 \, r_{lc}^{-2} \, r_{ts}^{-1} 
 \propto P^{-3/2} \, \dot{P}^{1/2}  \propto \dot{E}(t)^{1/2}$.
Therefore $B_{ts} \propto \dot{E}(t)^{1/2}$ was much higher at earlier times when the pulsar spin-down
was greatest.
Past the termination shock the field continues to decrease, though likely at a slower
rate than $B \propto 1/r$, since this would put the field at $\ll 1 \, \mu \rm{G}$ in the outermost
reaches of the nebula. Instead, we adopt a behavior of: 
\begin{eqnarray}
B(r,t) = 400 \, \left( \frac{r}{0.03 \, \rm{pc}} \right)^{\beta} \left( \frac{\dot{E}(t)}{2.8\times 10^{36}} \right)^{1/2} \, \mu \rm{G}
\end{eqnarray}
past the termination shock, where $\beta$ is a fit parameter and the 
$400 \, \mu \rm{G}$ coefficient is the approximate value at the termination shock. 
We should emphasize that this magnetic field model does not conserve magnetic flux,
since we assume that some magnetic field may be generated in the post-shock flow.  
The mean magnetic field within the boundaries of each zone is determined via a volume
integration of Equation 20, and 
this is the value used at each time step since each zone is treated as homogeneous in the SED code.  

\subsection{Diffusion}

A certain fraction of the injected particles will be lost from the nebula due to diffusive escape.
For each zone we compute the diffusive flux between zones by solving:
\begin{eqnarray}
J_i(E_e,t) = -D \, \frac{\partial n_i(E_e,t)}{\partial R} \, \, \, \rm{cm^{-2} \, s^{-1} ,}
\end{eqnarray}
with $J_i(E_e,t)$ the diffusive flux in particles per $\rm{cm^2} \, s^{-1}$ for zone $i$, 
$n_i(E_e,t)$, the particle density, and $D$ the diffusion coefficient.  The concentration gradient 
$\partial n_i(E_e,t)/ \partial R$ is evaluated as the concentration difference between adjacent
zones divided by the distance between zone midpoints.  
Solving Equation 21 for each zone at every time step gives the number of particles diffusing 
between zones.  
Hence, time permitting particles typically diffuse sequentially from inner zones to outer zones,  
stepping from Zone 1 to Zone 12.  The PWN-SNR interface will likely trap the majority of high-energy
particles within the PWN.  In addition, the surrounding SNR is expected to contain significant numbers
of high energy particles.  We therefore expect that particles will diffuse rather slowly out of the 
PWN and into the SNR, and so we arbitrarily set the density in the surrounding
SNR $n_{SNR}(E_e,t) = 0.8 \times n_{12}(E_e,t)$, which allows particles to slowly escape the PWN from Zone 12.  
The final SED depends very weakly on $n_{SNR}(E_e,t)$, however,
given the high degree of cooling and large final size of the nebula.  

Toroidal magnetic field lines tend to contain particles very effectively, and even with
relatively rapid Bohm diffusion
$\tau_{esc} \approx 40,000 \, (R / 10 \, \rm{pc})^2 (E_e/100 \, \rm{TeV})^{-1} \, (B/10 \, \mu \rm{G}) \, \rm{year}$
\citep{dejager+datai08}.
In the Bohm limit the cross-field mean free path $\lambda$
equals the particle gyroradius, which is exceedingly small:
$\lambda = 0.01 (E_e/100 \, \rm{TeV}) \, (B/10 \, \mu \rm{G})^{-1}$ parsec.  
If the field line structures contain radial components, however, particles diffuse
far more rapidly.  Turbulence and mixing caused by the passage of the reverse shock 
might provide the necessary disruption to the magnetic field structure to achieve radial field components.
Even in young PWNe where
significant reverse shock interaction has likely not yet occurred, however,
the simple model of purely toroidal magnetic field has been challenged by comparing
X-ray observations of PWNe with the predictions of the \citet{kc84b} model.  
Chandra observations of 3C 58 \citep{slane04} allowed comparison of the radial X-ray
spectral index variation with the predictions of \citet{kc84b}.  For any reasonable
values of $\sigma$ these authors found the observed photon index increased much slower
in 3C 58 than expected from the model of \citet{kc84b}.  
A similar deviation is observed for G$21.5 -0.9$ 
\citep{slane00}, implying that the magnetic field structure deviates from
a purely toroidal structure.    

The characteristic time scale for particle diffusion is determined by both the 
size of the nebula as well as diffusion coefficient $D$, such that
$\tau_{esc} \approx R^2/6D$.  Alternately,
the mean free path of a particle
$\lambda \equiv 3 D / c$.  
The electron diffusion coefficient is frequently
parameterized in power-law form $D(E_e) = D_0(E_e/E_{e,0})^{\delta}$.  For Bohm
diffusion $\delta = 1$, while
energy independent diffusion of course corresponds to $\delta = 0$.
To account for diffusion in a non-toroidal magnetic field
we select a Bohm-type diffusion with $\delta = 1$, such that mean free path and escape time scale as:
\begin{eqnarray}
\lambda & = & a \,  E_e  \nonumber \\
\tau_{esc} & = & R^2 (2 a c E_e)^{-1}
\end{eqnarray}
with $a$ left as a free parameter in the model such that the mean free path is allowed to vary 
from the Bohm value to account for complex nebular field structure.
For simplicity we ignore any perturbation the magnetic field
might have on the escape timescale, since clearly particles must primarily move along rather than
across magnetic field lines.

\subsection{Photon Fields}

Common practice in PWN SED modeling has been to adjust the dust IR 
energy density to maximize the agreement between model and data.  We instead fix the 
photon densities at published values, reducing the number of model fit parameters.
Adopting three primary photon fields 
(CMBR, far IR (dust), and starlight), we   
employ the interstellar radiation mapcube within the GALPROP suite \citep{porteretal05} 
to estimate the photon fields at the Galactic radius of PSR\,
J1826$-$1334. 
A distance of 
4 kpc in the direction of PSR J1826$-$1334 corresponds to 
a Galactic radius of 4.7~kpc; at this radius, dust IR photons
typically peak at $T \approx 32$ K with a density of $\approx 1$
eV$\rm \, cm^{-3}$, while stellar photons peak at $T \approx 2500$ K
with a density of $\approx 4$ eV$\rm \, cm^{-3}$.  At these densities IR and CMB photons
contribute approximately equally to TeV inverse Compton emission, while stellar photons are 
inconsequential.  
Synchrotron self-Compton emission is negligible for the magnetic fields present in
HESS J1825$-$137.

\subsection{SED Model}
A number of recent papers have applied various models to investigate the broadband emission from PWNe
(\citet{lemiereetal09}, \citet{gelfandetal09}, \citet{bucci10}, \citet{fang+zhang10}, \citet{slane10}, \citet{tanaka+takahara10}). 
\citet{gelfandetal09}, in particular, 
constructed a detailed 1-D dynamical model of PWNe magnetic field, pressure, size, energetics and 
spectral evolution inside a host SNR.  These authors demonstrated their model by fixing parameters to 
mimic the Crab Nebula.  While this model certainly helps shed light on compact PWNe such as the 
Crab, the markedly different sizes of X-ray and TeV emission regions in HESS J1825$-$137 (along with most 
other middle aged Vela-like PWN) pose a challenge to 1-D models.  

We therefore
treat the nebula as a series of twelve expanding bubbles in order to match the twelve H.E.S.S. spatial regions.
Each of these bubbles is treated as homogeneous, and filled
with particles (assumed to be electrons) as well as magnetic field.  
We compute SEDs from evolving the electron populations over various lifetimes in a series
of time steps, extending the single-zone approach taken in \citep{velax}, \citep{pwncat}, \citep{1825fermi}.

The twelve-zone modeling approach  permits investigation of the twelve 
H.E.S.S. spectral extraction regions, yet the number of spatial zones need
not be limited to the number of data regions and in theory one could utilize any
number of zones and then project those zones onto the sky for SED construction.  
The spatial grid size is completely arbitrary, and hence the
final SED plot should be independent of grid size. A finer grid of 
zones vastly increases computation time, however, and provides little new information given
the limited number of data regions.    
Nevertheless, as a consistency check we double the number of zones
(to 24) and plot the resultant SED.  As expected, the model curves change imperceptibly in all
but the innermost region.  The slight alteration in Region 1 is due to the steep variation in the 
magnetic field near the pulsar.    
This lack of dependence of the SED on the spatial zone size provides a check of
the SED model.

\subsection{Model Implementation}

Incorporating diffusion and projection effects between zones necessitates that all zones evolve 
in parallel, rather than independently. Evolution begins with Zone 12 at $t=t_{i,12}=0$, with escaping 
particles simply lost from the nebula.  
Once injection ends for Zone 12, at $t_{i,11}$, 
Zone 11 begins to evolve, with its escaping particles injected into Zone 12.  Once the injection 
period for Zone 11 ends, Zone 10 begins the injection phase, with Zones 11 and 12 in the cooling 
phases.  This process is repeated until the current age of the pulsar is reached, with all twelve 
zones evolving simultaneously.   Figure 5 shows a schematic of the evolution of the nebula for 
three spatial zones.

\begin{figure}[tbp!]
\epsscale{1.2}
\plotone{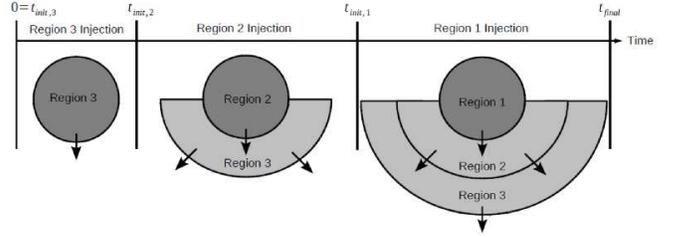}
\caption{
Snapshots of the evolution of nebular bubbles over three injection epochs.  
Dark gray corresponds to injection from the pulsar,
light gray denotes the cooling phase, and arrows indicate diffusing particles.  
}
\end{figure}

To compute the PWN SED, injected electrons are evolved through both the injection and
cooling phases of each bubble's lifetime, with the age of each phase computed via Equation 13.
Each zone shares the same injection spectrum, modulated by the spin down power.  
Cooling is computed from synchrotron 
\citep[Equation 4.5]{bg70}, inverse Compton \citep[Equation 2.56]{bg70}, and 
adiabatic losses (Equation 17).  
Evolution occurs in a series of time steps,
such that during the injection phase at each time step we perform the following procedures:

\begin{enumerate}
\item Calculate the size of the zone by integrating Equation 16

\item Compute the mean magnetic field by integrating over the volume of the zone via Equation 20

\item Inject pulsar wind particles into the zone according to Equation 10 

\item  Inject particles diffusing from adjacent zones according to Equation 21 (if including diffusion)

\item  At each particle energy $E$, compute $\delta E$  from synchrotron, inverse Compton, and adiabatic losses 

\item  Calculate the subsequent particle spectrum at time $t+\delta t$: $E(t+\delta t) = E - \delta E$

\item  Remove escaping particles according to Equation 21 (if including diffusion)
\end{enumerate}
Once injection ends, cooling is computed for the remainder of the evolution, following the same steps above
(save for step 3).  At each time step this process is repeated for all zones 
with $t_i > t$. When evolution completes, 
the synchrotron \citep[Equation 8.58]{longair}, 
and IC \citep[Equation 2.48]{bg70} fluxes are computed from the final
electron spectrum and magnetic field for each zone.  Finally, the flux within each region
is tabulated using the coefficients of Table 5.  

Model fitting is achieved by minimizing the $\chi^{2}$ between model and data
using the downhill simplex method described in \citet{pressetal92}.
Data consists of both X-ray and H.E.S.S. spectral flux points (77 in total): 12 in Region 1, 11 in Region 2,
9 in Region 3, and 5 data points for all exterior regions.
For each ensemble of $N$ variable parameters we evolve the system over the pulsar 
lifetime and calculate $\chi^{2}$ between model curves and flux data points.  
The simplex routine subsequently varies the parameters of interest
to minimize the fit statistic.  
We estimate parameter errors by computing $\chi^{2}$ for a sampling of 
points near the best fit values and using these points to fit the $N-$dimensional ellipsoid 
describing the surface of $\Delta \chi^{2} = 2.71$.  Under the assumption of Gaussian errors, 
the projected size of this $\Delta \chi^{2} = 2.71$ ellipsoid onto each parameter axis defines
the 90\%
multi-parameter (projected) error.  

\section{Results}

\subsection{Uniform Expansion, Constant Magnetic Field (Model 1)}

We begin by considering a very simple model of constant magnetic field ($\sim 3 \, \mu$G), 
uniform expansion ($v(r,t)= \rm{constant}$), and ignore diffusive effects.  
Uniform expansion implies that each zone 
undergoes expansion for $1/12^{\rm{th}}$ of the pulsar lifetime.  
Injection begins at pulsar birth ($t=0$) for Zone 12, while
Zone 1 begins evolution at $T - T/12$, where
$T$ is the current age of the pulsar.
The large size of the PWN implies a
relatively high age, and the braking index and $P_0$ selected in Table 6
give a total age of 21 kyr.
This corresponds to a mean expansion velocity of $3900 \, \rm{km \, s^{-1}}$.
We initially adopt a minimum particle energy $E_{min}=100$ GeV, 
and electron injection fraction $\eta = 0.8$, which is a reasonable value
for the expected particle dominated wind.  

\begin{figure}[tbp!]
\begin{center}
\end{center}
\epsscale{1.2}
\plotone{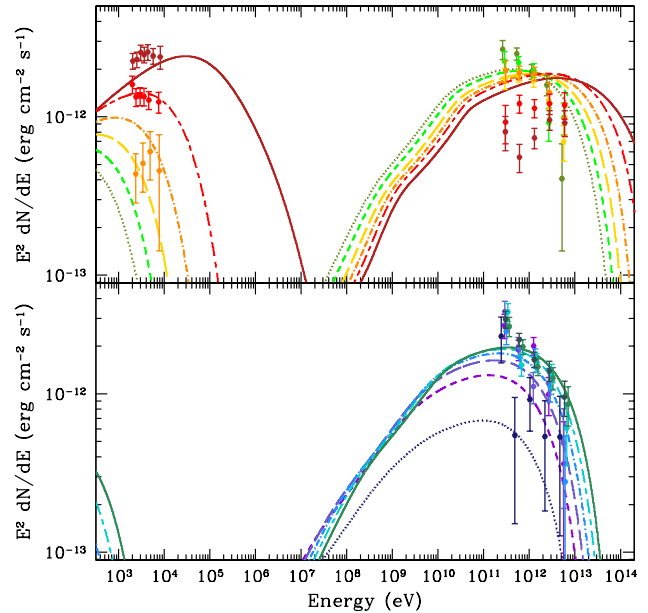}
\caption{\label{fig:SED1}
Model 1. Fitted simple spectral energy distribution model of the nebula with a constant magnetic field 
($2.9 \, \mu$G)
throughout the nebula for all time, constant expansion velocity such that each zone undergoes 
injection for $1/12^{\rm{th}}$ of the pulsar age (21 kyr), and $\chi^2/d.o.f. =  575/74$. 
The SED plot is split into two panels for clarity, with a color scale evolving from dark red 
(Region 1, youngest), to green, to blue, to violet (Region 12, oldest).  
The line type evolves as well, from
solid (inner, youngest) to long dot-dashed, short dot-dashed, long dashed, short dashed, 
and finally dotted (outer, oldest).  
The upper panel shows the model curves 
and H.E.S.S. data for the inner six regions, and \emph{Suzaku} data for the inner three regions, while
the lower panel shows the outer six regions.  
The fit is clearly quite poor, requiring fewer particles in the inner regions (and a higher
magnetic field), with less cooling 
and more power in the outer regions.  The gradual decrease of the cooling break energy with 
increasing region is primarily due to energy dependent synchrotron losses. 
}
\end{figure}

The model gives a poor match to the data,
apparent in the SED plot of Figure~\ref{fig:SED1}.  Applying the full fitting machinery to 
five variables (magnetic field $B$, initial spin period $P_0$, braking index $n$, electron
power-law index $p$, electron cutoff $E_{cut}$) yields 
an extremely low initial spin period $<1$ ms.    
Figure~\ref{fig:SED1} 
therefore fixes the braking index ($n=3$) and initial spin period ($P_0 = 15$ ms) to reasonable
values; the other three variables are fit and 90\% multi-parameter error estimates are computed, as
shown in Table 6.  The fit is still quite poor, and the fit value of $E_{cut} = 3200$ TeV is
implausibly large. Inspection of Figure~\ref{fig:SED1} indicates that
the normalization of each individual zone is highly skewed.  The innermost zone 
has far too much VHE flux, while Zone 12 at the nebular frontier 
has far too little VHE flux.   

\subsection{Velocity Profile, Constant Magnetic Field (Model 2)}

\begin{figure}[tpb!]
\begin{center}
\end{center}
\epsscale{1.2}
\plotone{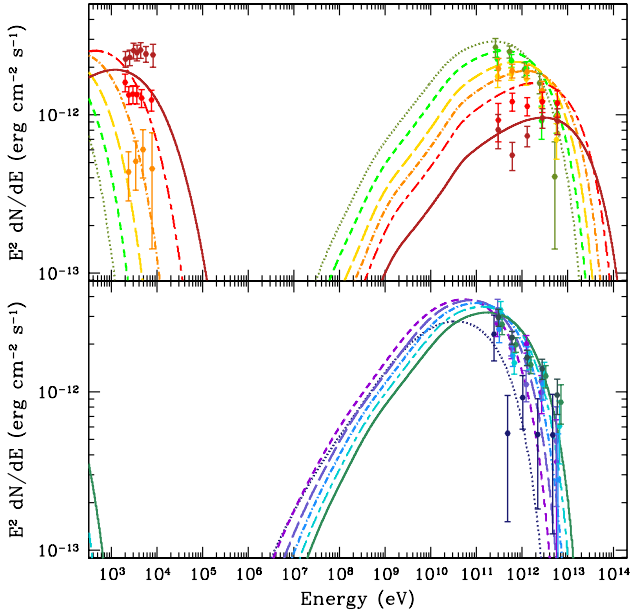}
\caption{\label{fig:SED2}
Model 2. Fitted SED model with a constant $5.6 \, \mu$G magnetic field
throughout the nebula for all time, $v(r) \propto r^{-0.36}$ such that the length of the
injection phase increases with zone number, and $\chi^2/d.o.f. = 236/71$.
The model curve line scheme is described in Figure~\ref{fig:SED1}.  
The radial velocity profile improves the fit over Figure~\ref{fig:SED1}, 
though the poor fit to the innermost X-ray
data and excessive cooling in the outer zones suggest a radial magnetic field profile may
provide a better fit.
}
\end{figure}

Though each region occupies the same $6\arcmin$ width, the normalization issues of 
Figure~\ref{fig:SED1} intimate that 
the injection time of the zones must vary with distance from the pulsar, and hence
the flow speed in the nebula must not be constant.  
We therefore adopt a velocity profile in the nebula of
$v(r,t) = r^{\alpha} \, C(t)$ (Equation 16).   
The best fit with this model and six variables is shown in Table 7.
The radial velocity profile results in a lengthening of the injection phase 
with increasing radial distance from the pulsar, and the values $n$ and $P_0$ 
yield an age of 28 kyr.  
This model is shown in Figure~\ref{fig:SED2}, is far superior to 
Figure~\ref{fig:SED1}, and reproduces the VHE steepening trend 
moderately well for the inner six regions. 
Yet the outer model curves 
 are all far steeper than the VHE data, implying an excess of
synchrotron cooling for the $\sim 40$ TeV particles required to upscatter CMB
photons to the $\sim 5$ TeV highest energy H.E.S.S. data point (Equation 4).  
As a result, in the outermost regions
(presumably corresponding to the oldest electrons), electrons of energy $\sim 40$ TeV
have almost completely burned off.  
In the innermost region the X-ray flux is significantly under-produced, suggesting the field
must be $> 5 \, \mu \rm{G}$ near the pulsar.  
Evidently a non-uniform magnetic field profile is required, with higher 
field in the interior to match X-ray fluxes and lower field in the exterior 
in order to reduce synchrotron cooling for the oldest electrons.

\begin{figure}[tpb!]
\begin{center}
\end{center}
\epsscale{1.2}
\plotone{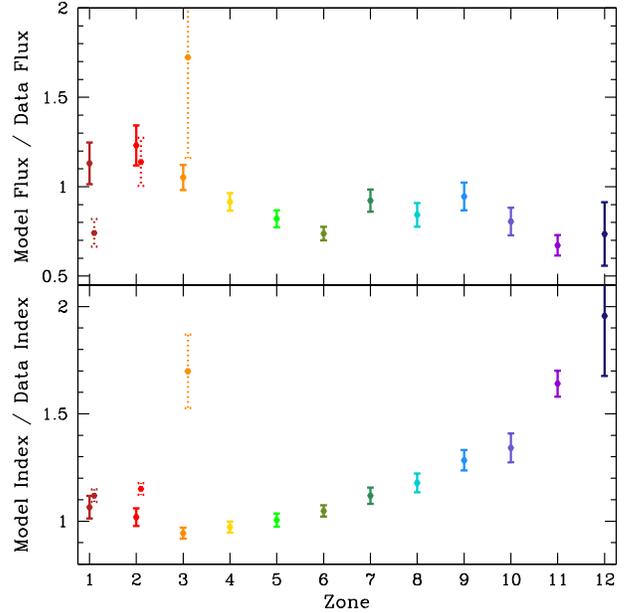}
\caption{\label{fig:ratio2}
Model 2. {\bf Top:} Ratio of model flux to the data listed in Tables 3 and 4, 
with X-ray $1-8$ keV flux ratios for the inner three zones (dotted) and 
VHE $0.25-10$ TeV flux ratios (solid) for all twelve zones.  
{\bf Bottom:} Power-law
indices over the same energy bands, with the model steadily worsening in the the outer zones.  
}
\end{figure}

In order to better visualize the results of the SED fitting, 
we also show the ratios between the model zone fluxes and spectral
indices and the measured data values (Figure~\ref{fig:ratio2}). 
The model flux is computed by integrating the flux in the desired energy band 
(either $1-8$ keV or $0.25-10$ TeV) and model indices are computed by fitting to a power-law 
in the appropriate band.  The model fluxes and indices are compared to the data from Tables 3 and 4.

\subsection{Velocity Profile, Magnetic Field Profile (Model 3)}

\begin{figure}[tpb!]
\begin{center}
\end{center}
\epsscale{1.2}
\plotone{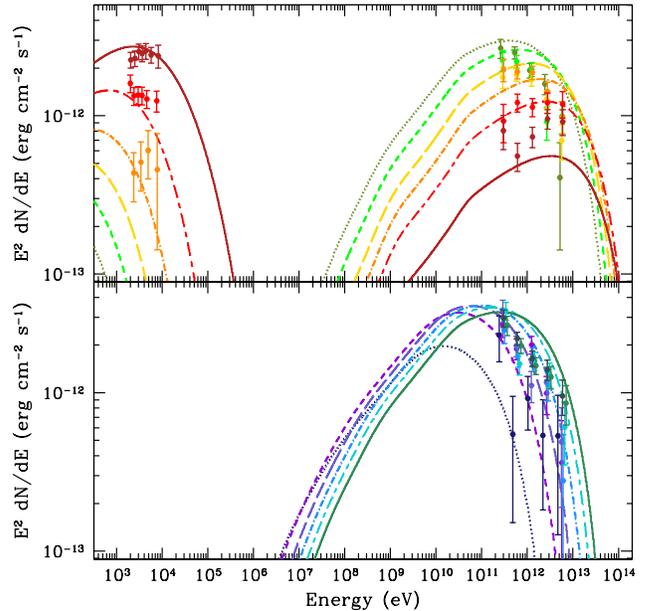}
\caption{\label{fig:SED3}
Model 3. Fitted SED model with 
$B(r,t) \propto r^{-0.74} \, \dot{E}(t)^{1/2}$,
 $v(r) \propto r^{-0.60}$, and $\chi^2/d.o.f. = 217/71$. 
The model curve line scheme is described in Figure~\ref{fig:SED1}.  
The radial magnetic field profile results in a superior fit to the innermost
X-ray data compared to Figure~\ref{fig:SED2}, 
though we still suffer from excessive cooling losses in the outermost zones.
}
\end{figure}

\begin{figure}[tpb!]
\begin{center}
\end{center}
\epsscale{1.2}
\plotone{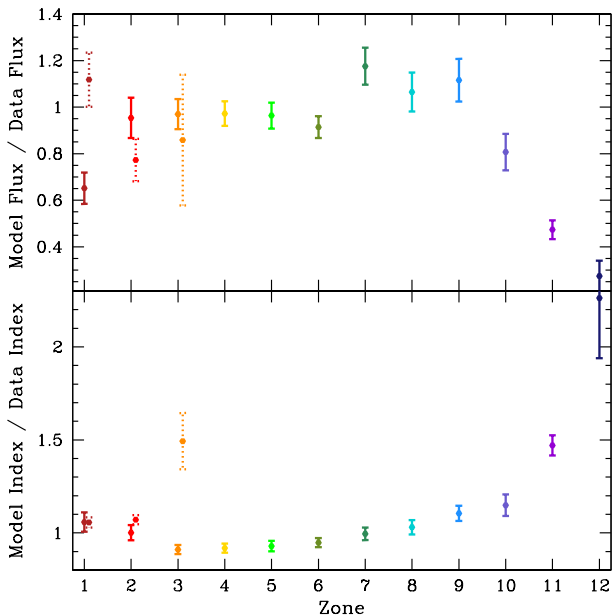}
\caption{\label{fig:ratio3}
Model 3. {\bf Top:} Ratio of model flux to the data listed in Tables 3 and 4, 
with X-ray $1-8$ keV flux ratios for the inner three zones (dotted) and 
VHE $0.25-10$ TeV flux ratios (solid) for all twelve zones. The X-ray flux ratios are much
improved over Figure~\ref{fig:ratio2}.
{\bf Bottom:} Power-law
indices, with the model steadily worsening in the the outer zones.  
}
\end{figure}

In hopes of improving the fit of Figure~\ref{fig:SED2}, 
we invoke a radial magnetic field profile of the form
$B(r,t) \propto r^{\beta} \, \dot{E}(t)^{1/2}$
(Equation 20), such that 
as particles traverse the nebula they experience an ever decreasing magnetic field.  
Similar to section 4.2, we select six fit parameters, though instead of 
fitting the final magnetic field value we fit the magnetic profile index $\beta$.
The best fit model yields an age of 34 kyr, and Table 6 reveals an improved 
fit statistic over the constant magnetic field model.
The non-constant magnetic field does a much better job of matching the innermost
X-ray data (see Figures~\ref{fig:SED3} and~\ref{fig:ratio3}  ), though 
 we still have difficulty matching the VHE indices 
in the outer (purple) zones.  

\newpage

\subsection{Velocity Profile, Magnetic Field Profile, Diffusion (Model 4)}

The severe under-prediction of VHE flux and over-prediction of VHE index
in the outer two zones
 of Figures~\ref{fig:SED3} and \ref{fig:ratio3} 
indicates a general lack of electrons with $E > 10$ TeV.  
One possibility of addressing this issue is simply reduced cooling by lowering the magnetic field. 
Yet even in Figures~\ref{fig:SED2} and \ref{fig:ratio2} with a low $6 \, \mu$G field 
the  Region 12 model is far steeper than the data due to excessive cooling.  
In addition, Figures~\ref{fig:SED3}  and \ref{fig:ratio3} demonstrate
that higher fields are needed in the inner zones, which only serves to hasten
synchrotron burn-off.  Particle re-acceleration as particles traverse the nebula might explain the 
VHE flux $\sim 80$ pc from the pulsar, perhaps due to reverse shock interactions.  Such acceleration is highly speculative,
however, and beyond the scope of this paper.  
A more standard method of maintaining the high energy particle population far from the pulsar, 
which we explore below, is rapid energy-dependent diffusion with mean free path
$\lambda \propto E_e$.

\begin{figure}[tpb!]
\begin{center}
\end{center}
\epsscale{1.2}
\plotone{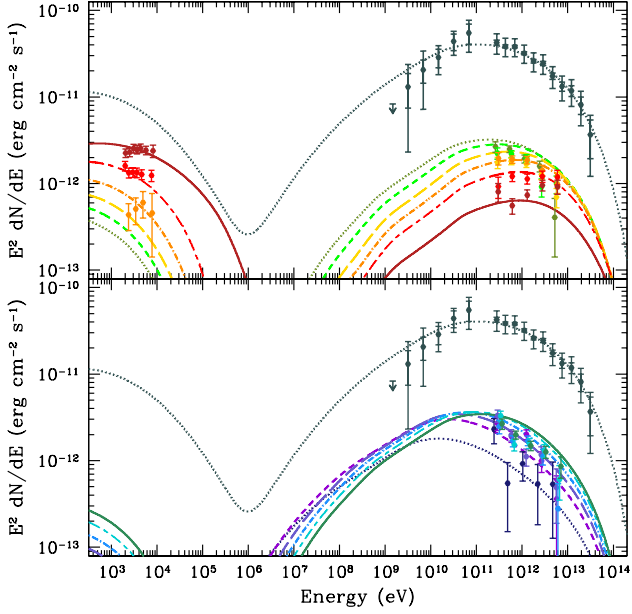}
\caption{\label{fig:SED4}
Model 4. Fitted SED model, scaled to show the aggregate data, with 
$B(r,t) \propto r^{-0.69} \, \dot{E}(t)^{1/2}$,
$v(r) \propto r^{-0.51}$, 
$\lambda = 1.8 \, (E_e/100 \, \rm{TeV})$ parsec, and $\chi^2/d.o.f.=119/70$.
The model curve line scheme is described in Figure~\ref{fig:SED1}.  
The inclusion of rapid diffusion among zones provides enough high energy electrons in the outer
nebula to match the H.E.S.S. data in Zones $10-12$. 
The slate data points denote
the aggregate \emph{Fermi} LAT and H.E.S.S. data, with both statistical and systematic error
estimates.  The slate dotted line (representing the sum of flux from all twelve 
zones) is scaled up by 25\%
to account for the greater total flux in the aggregate extraction region (see Section 2.2).
The overall fit to the aggregate data with this scaling is quite impressive.   
}
\end{figure}

\begin{figure}[htpb!]
\begin{center}
\end{center}
\epsscale{1.2}
\plotone{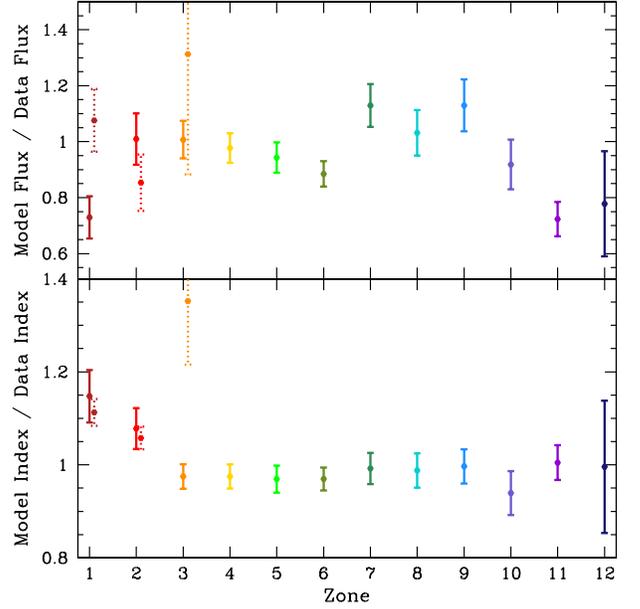}
\caption{\label{fig:ratio4}
Model 4. {\bf Top:} Ratio of model flux to the data listed in Tables 3 and 4, 
with X-ray $1-8$ keV flux ratios for the inner three zones (dotted) and 
VHE $0.25-10$ TeV flux ratios (solid) for all twelve zones.
{\bf Bottom:} Power-law
indices. For both upper and lower panels the outer-zone 
ratios are far superior to Figures~\ref{fig:ratio2} and \ref{fig:ratio3}.
}
\end{figure}

For this model we select seven variables:
velocity radial profile index $\alpha$, magnetic profile index $\beta$, initial spin
period $P_0$, braking index $n$, electron power-law index $p$, 
electron exponential cutoff $E_{cut}$, 
and mean free path $\lambda$. 
Figure~\ref{fig:SED4} shows the SED of the best fit model, 
scaled to display the aggregate gamma-ray data.  This aggregate data is discussed in 
Section 5.  Figure~\ref{fig:ratio4} displays the flux and index ratios for this model.  
The overall fit to the twelve spectral regions is far superior to previous models,
 particularly in the outer nebula.  The largest departure of the model from the data
is the X-ray spectral index of Region 3, though we remind the reader that the 
very hard Region 3 X-ray index reported in Table 4 may reflect residual background contamination. 
Table 6 lists the fit parameters, while Table 7 lists derived parameters from the fit
(injection time, final magnetic field, and $\chi^2$). As expected from the scaling relations
 of section 3.1, the particle residence time of Zone 1 is well under 1000 years.   
The braking index $n$ and initial spin period $P_0$ yield an age of 
 $40 \pm 9$ kyr and a mean expansion velocity of $2000 \, \rm{km \, s^{-1}}$. 
The mean free path of $\lambda = 1.8 (E_e/100 \, \rm{TeV})$ parsec predicts an escape timescale of
$\tau_{esc} \approx 90 \, (R / 10 \, \rm{pc})^2 (E_e/100 \, \rm{TeV})^{-1} \, \rm{year}$. 
This model does have some difficulty matching the H.E.S.S. flux and slope in Region 1, which
is due to the rapid diffusive loss of high energy particles.  Most likely, the simple
diffusive model employed vastly oversimplifies the complex nature of this source, and diffusion
coefficients vary in both space and time.  Nevertheless, the $\chi^2/d.o.f.$ for Model 4 indicates
that the general fit is quite good.  

As a check of our magnetic field model, we also allow the normalization at the termination shock to
vary from its adopted value of $400 \, \mu \rm{G}$.  
The best fit model with this extra fit parameter yields a 
termination shock field of $320 \, \mu \rm{G}$ 
with no improvement of the fit statistic ($\chi^2/d.o.f. = 118/69$,
versus $\chi^2/d.o.f. = 119/70$ for the field fixed.  
Fitting the magnetic field
normalization therefore does not yield a markedly better fit, and the termination shock
value of $400 \, \mu \rm{G}$  appears quite reasonable. One further check of the magnetic field model
is achieved by fitting a model with diffusion over uniform magnetic field (simply replacing
the magnetic field radial index of Model 4 with a constant magnetic field value).  
This model cannot adequately match the data, 
yielding a best fit of $\chi^2/d.o.f. = 429/70$,  and rules out uniform magnetic field as a 
viable option.  
The adopted IR photon density of $\approx 1$ eV$\rm \, cm^{-3}$ 
could easily vary by a factor of $\sim 2$, 
though varying the photon density does not qualitatively change the results presented
here.  A higher (lower) IR photon density means fewer (more) high energy electron are required
to match the HESS data, implying a slightly softer (harder) electron spectral index or lower (higher)
high energy cutoff.

\begin{table*}[tpb!]
\caption{Multiwavelength SED Fit To All Regions}
\centering
\begin{tabularx}{1.0\textwidth}{cZZZZZZZc}
\hline\hline
Model $\T \B$ & $\alpha \; (v \propto r^{\alpha})$ & $B / \beta \, (B \propto r^{\beta})$ & $P_0$ (ms) & $n \; (\dot{\Omega} \propto \Omega^n) $ & $p $ & $E_{cut}$ (TeV) & $\lambda \; (\frac{E_e}{100 \,{\rm TeV}}) \, {\rm pc}$ & $\chi^2$/d.o.f. \\
\hline
$1 \T \B$  & $-$ & $2.9 \pm 0.3 \tablenotemark{a}$ & $15\tablenotemark{b}$ & $3\tablenotemark{b}$ & $2.42 \pm 0.06$ & $3800 \pm 800$ & $-$ & $575/74$ \\
$2 \T \B$  & $-0.36\pm0.03$ & $5.6\pm0.4\tablenotemark{a}$ & $12\pm3$ & $2.5\pm0.1$ & $2.30\pm0.01$ & $130\pm10$ & $-$ & $236/71$ \\
$3 \T \B$  & $-0.60\pm0.08$ & $-0.74\pm0.02$ & $14\pm7$ & $2.2\pm0.1$ & $2.31\pm0.02$ & $190\pm30$ & $-$ & $217/71$ \\
$4 \T \B$  & $-0.51 \pm 0.09$ & $-0.69 \pm 0.01$ & $13 \pm 7$ & $1.9 \pm 0.2$ & $2.24 \pm 0.02$ & $230 \pm 90$ & $1.8 \pm 0.3$ & $119/70$ \\
\hline
\end{tabularx}
\centering 
\tablenotetext{a}{ Final (uniform) magnetic field ($\mu$G)}
\tablenotetext{b}{ Held fixed}
\end{table*}
\normalsize

\begin{table}[htpb!]
\centering
\caption{Derived Parameters for Model 4}
\begin{tabular}{c@{\hspace{0.5cm}}c@{\hspace{0.5cm}}c@{\hspace{0.5cm}}c@{\hspace{0.5cm}}c@{\hspace{0.5cm}}c}
\hline \hline
\centering
Zone  \T \B & $R$ (pc) & $\delta t_i \,(\rm{year})\tablenotemark{a}$ & $B \, (\mu \rm{G})\tablenotemark{b}$ & $\chi^2$ & $\#$
                                                                         data points\\
\hline
$1$ \T & $7$ & $600$ 	     & $12$   & $23$  & $ 12$ \\
$2$ & $14$ &  $1200$	     & $6.6$   & $13$  & $ 11$ \\
$3$ & $21$ &  $1500$	     & $4.8$   & $6.6$  & $ 9$ \\
$4$ & $28$ &  $1900$ 	     & $3.8$   & $4.6$  & $ 5$ \\
$5$ & $35$ &  $2200$ 	     & $3.2$   & $8.0$  & $ 5$ \\
$6$ & $42$ &  $2500$ 	     & $2.8$   & $10$  & $ 5$ \\
$7$ & $49$ &  $2900$ 	     & $2.5$   & $8.9$  & $ 5$ \\
$8$ & $56$ &  $3300$ 	     & $2.2$   & $11$  & $ 5$ \\
$9$ & $63$ &  $3800$ 	     & $2.1$   & $7.4$  & $ 5$ \\
$10$ & $70$ &  $4400$ 	     & $1.9$   & $6.4$  & $ 5$ \\
$11$ & $77$ &  $5500$ 	     & $1.8$   & $16$  & $ 5$ \\
$12$ & $84$ &  $9900$ 	     & $1.7$   & $4.4$  & $ 5$ \\
\hline
\end{tabular}
\centering
\tablenotetext{a}{ Injection timescale}
\tablenotetext{b}{ Final magnetic field} 
\end{table}
\normalsize

\subsubsection{Relativistic Maxwellian Plus Power-Law Injection Spectrum}

The SED model described in the previous section takes into account only X-ray and VHE H.E.S.S. data,
though ideally the model should be consistent with \emph{Fermi} LAT data as well.  
Note that while we do not fit to the LAT points in Figure~\ref{fig:SED4}, the
rough agreement between these data and the integrated (total) model flux
in the LAT band is encouraging.
However, the \emph{Fermi} LAT photon index of $\Gamma = 1.4 \pm 0.2$ \citep{1825fermi} predicts an electron index
of $p = 2 \Gamma - 1 = 1.8 \pm 0.4$.  This is somewhat harder than
the electron index of $p = 2.2$ utilized in Figures ~\ref{fig:SED4} and \ref{fig:ratio4}, 
and an injection spectrum with fewer low energy electrons
might plausibly improve the fit to the LAT data.  

One option is adopting the relativistic Maxwellian plus power-law
tail electron spectrum proposed by \citet{spitkovsky08}, and described in modeling terms
by \citet{1825fermi}.  
Fitting with this model requires one additional parameter (the temperature of the Maxwellian), 
 bringing the total 
up to eight.  
The best fit model with this injection spectrum yields a fit somewhat worse than the power-law case 
($\chi^2/d.o.f. = 123/69$, versus $122/70$ for the power-law spectrum).   
In addition, an extremely short initial spin period of $P_0 \sim 2$ ms is required,
and the low $kT = 0.2$ TeV 
results in a severe
over-prediction of GeV flux and a far worse fit to the LAT data than the power-law case.
\citet{1825fermi} found an adequate fit to the aggregate data with a one-zone model, yet
the inclusion of multiple zones seems to wash out any advantage of a relativistic Maxwellian electron
spectrum.

\newpage

\section{Discussion}

As an initial check of the validity of the scalings applied to both the \emph{Suzaku}
and H.E.S.S. data, we refit Model 4 with the raw data.  Using the original
H.E.S.S. data points and X-ray data taken from regions coinciding with the HESS quarter annuli,
the best fit parameters are within $1\sigma$ of the values listed in Table 6
for Model 4, except for the magnetic field index.  The different H.E.S.S. flux
ratios between interior zones results in a magnetic field index that differs from the 
Model 4 value by +0.02, or $2\sigma$.  
Despite the similarity in fit parameters, however, the fit statistic 
increases dramatically, from $\chi^2/d.o.f. = 119/70$ for Model 4, to 
$\chi^2/d.o.f. = 169/70$.  Nearly all of this increase in $\chi^2$ 
comes from the inner two regions.
The H.E.S.S. excess map (Figure 1b) reveals 
that there appears to be significantly more flux $6\arcmin-12\arcmin$ 
from the pulsar than within $6\arcmin$ of the pulsar.  
Yet due to the small original H.E.S.S. extraction region for the $6\arcmin-12\arcmin$ 
region (which is 25\% 
smaller than the $0-6\arcmin$ region), the reported H.E.S.S. 
fluxes are identical for the inner two regions.  The use of identical fluxes 
for the inner two zones therefore results in a worse fit for the model, 
which assumes radial variations in the flow speed and a greater 
number of particles farther from the pulsar. Nevertheless, the commonality of the fit
parameters with the raw and scaled data provides some support for the scaled
approach taken throughout this paper.  One further check of the X-ray scaling
is to fit the scaled H.E.S.S. data and raw, unscaled X-ray data.
With this slightly lower X-ray flux the best fit parameters are well within
errors for Model 4, and give a nearly identical $\chi^2$. 

Of the four models scenarios described above, the final model which incorporates diffusion provides
by the far the best fit to the data. 
The relativistic Maxwellian injection spectrum yields a somewhat worse fit statistic to
the simple power-law case, and though
the power-law injection spectrum also better matches the \emph{Fermi} LAT data,
definitive discrimination between lepton injection spectra is not possible at this stage.
Yet we can convincingly argue that the pulsar wind must inject
significant numbers of electrons up to energies of at 
least 100 TeV, the low energy power-law index must be $p \sim 2$ in order to match the \emph{Fermi}
LAT data, that a high percentage of the pulsar spin-down energy must go into leptons, and that some
mechanism (we invoke diffusion) maintains high energy particles far from the pulsar.   
In the following discussion we consider the simple power-law spectrum injection spectrum (Model 4) 
shown in Figures ~\ref{fig:SED4} and \ref{fig:ratio4}.

The sum of fluxes from the twelve spatial zones has been scaled up some 25\% 
in Figure~\ref{fig:SED4} 
due to the fact that our spectral extraction regions do not capture the entirety of
the counts within the aggregate nebula.  This scaled flux ideally should
match the \emph{Fermi} LAT and H.E.S.S. aggregate data points (which both represent larger
spatial regions), which appears to 
be the case for Figure~\ref{fig:SED4}.
Computing the fit statistic for the aggregate H.E.S.S. and LAT data  
gives $\chi^2 = 27$ for
the 16 gamma-ray data points (assuming only statistical errors on the data points, and ignoring the LAT upper limit).
Given the very small statistical errors on the H.E.S.S. data points the agreement between
gamma-ray data and the summed model curves is remarkable.  
For comparison, the best fit model utilizing the relativistic Maxwellian injection spectrum nets
a $\chi^2 = 50$ for
the 16 gamma-ray data points.
The `missing' $\approx 25$\% 
of flux implies a missing population of particles contributing
to the IC flux of the nebula.  Such particles have likely diffused to the north and east of the pulsar
exterior to our extraction regions.
A tweak of the initial pulsar spin-down from 13 ms to 11 ms (well within errors)
provides $\approx 40$\%
greater spin-down energy, more than enough to
enough to raise the model bulk nebular flux to the observed level.

The models proposed
above utilize $\eta = 0.8$, and we find an adequate fit to the X-ray and H.E.S.S. data for $P_0 = 13$ ms.
Neglecting the `missing' energy component, this connotes a total of $1.1 \times 10^{50}$ erg of pulsar spin-down energy.
Disregarding losses, we therefore expect of order 
$10^{49}$ erg in magnetic field energy.  While the magnetic field
evolution and profile adopted in our model does not take into account the total energy budget 
of the pulsar, a reasonable model should nevertheless contain $\sim 10^{49}$ erg of
magnetic energy for the choice of $\eta=0.8$. 
Computing the magnetic field energy from the final magnetic field profile 
$B(r,T) = 400 (r/0.03 \, \rm{pc})^{-0.69} \, \mu \rm{G}$
yields an energy 
of $4.4 \times 10^{48}$ erg, 
which is quite reasonable given that the magnetic field likely
experiences some losses due to expansion.  
Even though the vast majority of the pulsar energy initially goes into particles, 
summing the energy content of the final electron spectrum reveals that electron cooling losses  
leave only $6.9 \times 10^{48}$ erg remaining in the form of leptons.  
We can also compute the value of $\sigma$ from the assumed termination shock radius of
$0.03$ pc and $400 \, \mu \rm{G}$ field strength.  Adopting a flow speed at the termination
shock of $c/3$ and a spherical geometry, one achieves an estimate of 
$7\times 10^{35} \, \rm{erg \, s^{-1}}$ of magnetic energy flux at the termination shock.  
For the pulsar spindown value of $\dot{E} = 2.8\times10^{36} \, \rm{erg \, s^{-1}}$ this 
yields $\sigma \approx 0.3$.  This value is consistent with the adopted value of $\eta = 0.8$,
which corresponds to a value of $\sigma = (1-\eta)/\eta = 0.25$.
One further consistency check for our model comprises comparing the available spin-down power to the total
energy present in the nebula plus cooling and escape losses.  Indeed, summing the leptonic energy,
magnetic energy, cooling losses, and escape losses yields $ 1.1 \times 10^{50}$ erg, precisely
the energy available from the pulsar.  

The energy lost due to adiabatic cooling goes into accelerating a shell of swept-up material, and 
we can use estimates of this swept-up mass as another check of the validity of the SED model.
For the adopted SNR density profile of a constant density inner core surrounded by a 
$\rho \propto r^{-9}$ outer envelope the transition between the two density regimes propagates out at
a constant velocity $v_t$, such that \citep{blondinetal01}: 
\begin{eqnarray}
v_t \approx 6500 \left( \frac{E_{SN}}{3 \times 10^{51} \, \rm{erg}} \right) ^{1/2}  
	\left( \frac{M_{ej}}{8 M_{\odot}} \right)^{-1/2} \rm{km \, s^{-1}.}
\end{eqnarray}
This is much faster than the Model 4 outer PWN expansion of $2000 \, \rm{km \, s^{-1}}$, 
so we can treat the PWN as always expanding into the 
constant density ejecta core, with density \citet{blondinetal01}:
\begin{eqnarray}
\rho_{ej}(t) & \approx & 3.7 \times 10^{-29}  \left( \frac{E_{SN}}{3 \times 10^{51} \, \rm{erg}} \right) ^{-3/2} 
	\left( \frac{M_{ej}}{8 M_{\odot}} \right)^{5/2} \nonumber \\
	& & \times \left( \frac{t}{20 \, \rm{kyr}} \right) ^{-3} \, \, \rm{g \, cm^{-3}.}
\end{eqnarray}
The amount of mass swept up by the PWN over its lifetime can be approximated as:
\begin{eqnarray}
M_{sw} = \int_{t_i}^T \rho_{ej}(t) \, \zeta 4 \pi r^2 dr ,
\end{eqnarray}
with $t_i$ the initial timestep, 
$r=v_{outer} t$ 
and $\zeta$ the filling factor of the PWN, equal to 0.31
for the adopted morphology.  
Integrating Equation 25 over the 40 kyr lifetime of the pulsar provides a very rough estimate of
$1 M_{\odot}$ of swept up mass.  
Accelerating $1 \, M_{\odot}$ to $2000 \, \rm{km \, s^{-1}}$ (the mean expansion velocity
for a pulsar age of $40$ kyr) requires some 
$4 \times 10^{49}$ erg.
For the parameters selected in Model 4, the 
SED model posits a total adiabatic loss of $3.4\times 10^{49}$ erg, quite close to the value quoted above.  
Recall that in the SED model adiabatic losses depend only on the
assumed PWN size evolution (Equation 17), and do not take into account any SNR properties.  Therefore
the fact that the two adiabatic loss estimates are not glaringly different
 provides some support to the soundness of the SED model.

The computed pressure within the entire nebula can be used as a further sanity check on our model by 
comparing to the expected PWN pressure after 40 kyr. 
Given the degree of particle cooling, the model's final particle pressure 
($P_{pwn} = \frac{E_p}{3 V_{pwn}} = 1.0 \times 10^{-13} \, \rm{erg \, cm^{-3}}$) 
is less than the magnetic pressure 
($\frac{B^2}{8 \pi} = 1.9 \times 10^{-13} \, \rm{erg \, cm^{-3}}$).  
Within a freely expanding SNR the PWN pressure
scales as \citep{vander01}:
\begin{eqnarray}
P_{pwn} \approx  3/25 \, \rho_{ej}(t) v_{pwn}^2, 
\end{eqnarray}
with $\rho_{ej}(t)$ given by Equation 24, and $v_{pwn}$ the
PWN expansion velocity.  After 40 kyr this predicts a mean nebular pressure of 
$P_{pwn} \approx 2 \times 10^{-13} \, \rm{erg \, cm^{-3}}$, in good agreement with the computed values.

\begin{figure}[tbp!]
\begin{center}
\end{center}
\epsscale{1.2}
\plotone{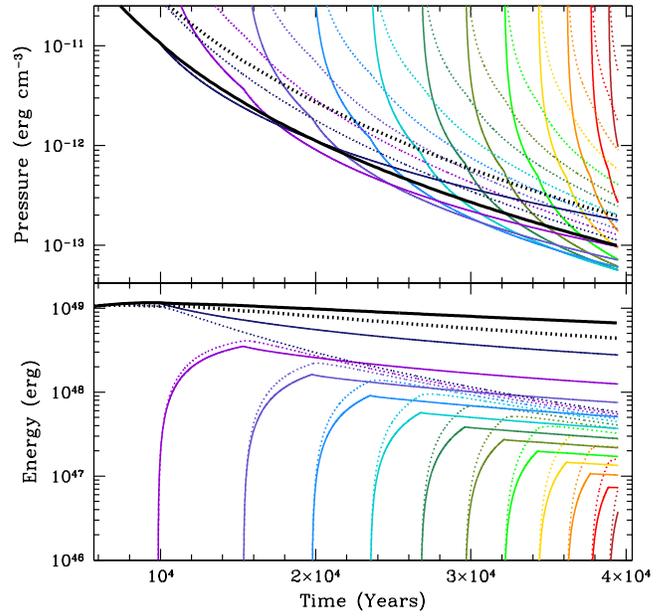}
\caption{\label{fig:p_en}
{\bf Top:} Magnetic field pressure (dotted) and particle pressure (solid) of each zone. 
Total pressures are marked by the thick black lines.  
{\bf Bottom:} Magnetic field energy (dotted) and particle energy (solid) of each zone.  
Black lines mark total energies.   
In both panels the values correspond to Model 4.
}
\end{figure}

The adopted termination shock radius of 0.03 pc predicts a pressure of 
$\approx 1 \times 10^{-9} \, \rm{erg \, cm^{-3}}$ at the termination shock (Equation 19), significantly
higher than the mean pressure of $\approx 3 \times 10^{-13} \, \rm{erg \, cm^{-3}}$.  
This high pressure at the shock rapidly dissipates away from the pulsar, however,
 with the bulk of the nebula
nearly isobaric at late times; each of the 10 outer zones has a pressure within
$\approx 50$\%
of the mean nebular pressure, as evidenced in the pressure and energy
plot of Figure~\ref{fig:p_en}.  This plot illustrates that the magnetic pressure
 exceeds particle pressure for the inner regions, and that
the total particle energy exceeds the bulk magnetic field energy in the nebula.   
The total magnetic field pressure scales as $P_{pwn,B} \propto t^{-3.0}$, while the total particle pressure
evolves as $P_{pwn,p} \propto t^{-2.9}$.  This is precisely the $P_{pwn} \propto t^{-3}$
evolution predicted by Equations 24 and 26 for a freely expanding PWN.

\begin{figure}[tbp!]
\begin{center}
\end{center}
\epsscale{1.2}
\plotone{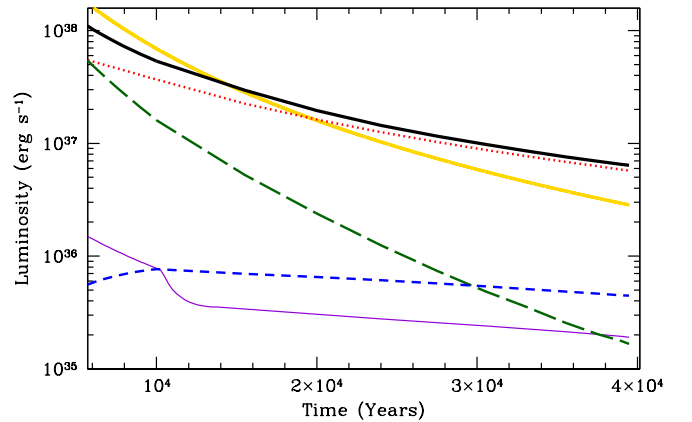}
\caption{\label{fig:lum}
Bulk luminosity of the nebula for Model 4.  The solid gold line denotes pulsar spin-down, while the 
solid black line shows the total luminosity from all processes.  Synchrotron (green long-dashed), 
IC (blue short-dashed), and adiabatic (red dotted) cooling is also shown.  
The violet line denotes diffusive escape losses from the nebula.
}
\end{figure}

In addition to the PWN pressure and energy we can also compute
 a variety of quantities which vary over the pulsar lifetime.  
We calculate the mean magnetic field evolution as
 $\overline{B} \propto t^{-1.6}$, which is quite similar to the $B \propto t^{-1.7}$
evolution computed by \citet{gelfandetal09} in their single-zone model during the free expansion phase. 
The various luminosities in the nebula are plotted in Figure~\ref{fig:lum},  
revealing that at late times the total luminosity of the nebula exceeds the pulsar spin-down power.  
This figure also demonstrates
that after a few thousand years adiabatic losses overtake synchrotron radiation to provide 
the dominant cooling mechanism, though particles of energy $E_e > 10$ TeV
primarily lose energy from synchrotron and IC cooling since for both the cooling 
timescale $\tau \propto E_e^{-1}$. Indeed, at late time the IC cooling actually exceeds
the synchrotron cooling due to the low mean magnetic field of $\approx 2 \, \mu \rm{G}$.

\begin{figure}[tbp!]
\begin{center}
\end{center}
\epsscale{1.2}
\plotone{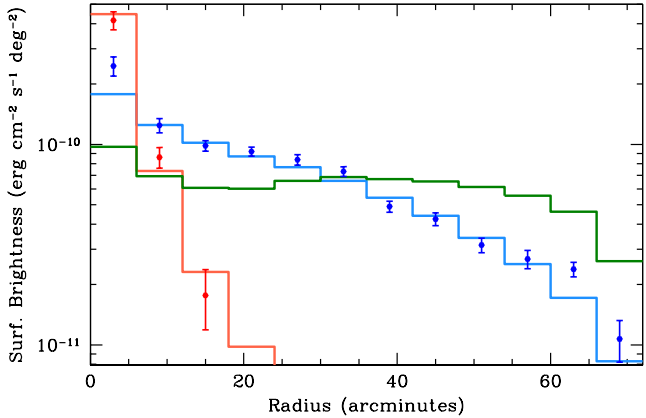}
\caption{\label{fig:mod_radial}
Nebula surface brightness profile indicating a remarkably uniform \emph{Fermi} LAT
profile.  Red data points denote \emph{Suzaku} X-ray data from Table 4
(statistical plus sytematic errors shown), while the 
red curve shows the Model 4 flux prediction in the 1-8 keV band.  Blue data points mark the H.E.S.S.
flux from Table 3, with the blue curve the 0.25-10 TeV model prediction.  The model 1-100 GeV band 
surface brightness is shown in green. 
}
\end{figure}

We construct radial flux profiles by summing the model fluxes in the \emph{Suzaku} and H.E.S.S.\ 
wavebands within each region, which can be compared with the data points plotted in Figure~2.  
We are also free to compute the profile in any arbitrary energy range, and so in 
Figure~\ref{fig:mod_radial}  we plot the projected surface brightness in the 
\emph{Fermi} LAT 1-100 GeV range.  Since the LAT energy range lies in 
the uncooled regime, the flux primarily tracks the energy content of electrons.  As a result,
the LAT flux peaks toward the outer realm of the nebula where the majority of electrons were injected
when the pulsar was much younger with a far higher $\dot{E}$,
and this gives that a rather
uniform counts map in the LAT energy band.

Comparison of point source and extended Gaussian morphologies in the $10-100$ GeV range
by \citet{1825fermi} yielded a significance of $\approx 8.5 \sigma$ for a source extension
of $0.56^{\circ} \pm 0.07^{\circ}$ centered $0.5^{\circ}$ southwest of the pulsar.  Similar results
are achieved with a uniform disk morphology, though the extension increases to 
$0.67^{\circ} \pm 0.02^{\circ}$. 
Fitting the source to a spatial template provided by the
H.E.S.S. excess map ($E > 200$ GeV) decreases the test statistic by $42 \approx 6.5 \sigma$, implying
that the LAT emission shows a distinctly different morphology from the H.E.S.S. emission.
Indeed, the LAT source extension of $\sim 0.6^{\circ}$ centered $0.5^{\circ}$ southwest of the pulsar is 
remarkably consistent with Figure~\ref{fig:mod_radial}.

\section{Conclusions}

Analysis of recently public \emph{Suzaku} X-ray data covering the majority of 
the very bright VHE source HESS J1825$-$137 indicate that X-rays extend only $\approx 1/4$ as far
as VHE gamma-rays.  The variable extent of the nebula in different wavebands, as well as the
observed VHE spectral softening away from the pulsar, is naturally explained by electron
cooling losses.  
Time-dependent 3-D modeling of the twelve data regions allows one to constrain the electron
injection properties, spin-down behavior of the pulsar, nebular velocity profile, and
magnetic field profile.  
A choice of $80$\% 
of the pulsar spin-down going into electrons is consistent with the data, 
though due to particle cooling at the current epoch
magnetic field energy is $\approx 1/2 \times$ the particle energy in the nebula.  
This cooling also results in a higher magnetic than particle pressure, with a nearly isobaric
particle pressure profile in the outer ten zones and the total pressure falling as 
$P_{pwn} \propto t^{-3}$.  For the best fit model (Model 4) an initial spin period of
$\approx 13$ ms provides adequate pulsar spin-down energy to power the nebula, and 
combined with the braking index of $n=1.9$ this predicts an age of $40$ kyr and 
an expansion velocity of $2000 \, \rm{km \, s^{-1}}$.  
The interior velocity profile is found to scale as $v(r)\sim r^{-0.5}$, while
the magnetic field profile scales as 
$B(r,t) \propto r^{-0.7} \, \dot{E}(t)^{1/2}$, with the field falling from
$\approx 400 \, \mu \rm{G}$ at the termination shock to $\approx 2 \, \mu \rm{G}$ in the 
far reaches of the nebula and the mean field scaling as $\overline{B} \propto t^{-1.6}$.

The \emph{Fermi} LAT  photon index of $1.4$ connotes an electron index near the canonical 2.0.
However, fitting to the X-ray/TeV and the higher energy portion of the electron spectrum, we
find a somewhat softer power law index of $p \approx 2.2$.  
A relativistic Maxwellian electron spectrum provides a somewhat worse fit to the X-ray
and H.E.S.S. data compared to a simple power-law injection, and the power-law case requires
fewer fit parameters and better matches the \emph{Fermi} LAT data.
Regardless of injection spectrum, the existence of 10 TeV photons in the 
outer reaches of the nebula 
requires substantial particle diffusion; adopting a diffusion timescale of
$\tau_{esc} \approx 90 \, (R / 10 \, \rm{pc})^2 (E_e/100 \, \rm{TeV})^{-1} \, \rm{year}$  
allows an adequate fit to the data.  This is in contrast to the common assumption of a purely toroidal
PWN magnetic field which could contain particles extremely efficiently. 
The model predicts a rather uniform \emph{Fermi} LAT surface brightness out to $\sim 1^{\circ}$ from
the pulsar, 
in good agreement with the recently discovered LAT source centered $0.5^{\circ}$ southwest of
PSR J1826$-$1334, with extension $0.6 \pm 0.1^{\circ}$. 

The multi-zone time-dependent SED model fitting approach applied in the paper extends simple
one-zone models to allow investigation of spatially dependent properties such as flow speed,
magnetic field, photon index, and flux levels.  The growing number of sources with spatially
resolved X-ray and VHE measurements (e.g. Vela-X, HESS J1303-631) are therefore prime targets
for such multi-zone modeling.  

\vspace{5 mm}

{\it Acknowledgments: } 
We thank Stefan Funk and Marianne Lemoine-Goumard for helpful discussions on PWN modeling
and \emph{Fermi} LAT data. We also would like to thank the referee for very 
thorough and helpful comments. 
This work was supported in part by the U.S. Department of Energy contract to SLAC no. DE-AC02-76SF00515
and by NASA grant NNX10AP65G.

\end{document}